\documentclass[fleqn,10pt]{wlscirep}
\usepackage[utf8]{inputenc}
\usepackage[T1]{fontenc}

\usepackage{balance,bm}
\usepackage{color, colortbl, xcolor}
\usepackage{url}
\usepackage{hyperref}
\hypersetup{
    colorlinks=true,
}
\usepackage{subcaption}
\usepackage{textcomp}
\usepackage{soul}
\usepackage{multirow}
\usepackage{wrapfig}
\usepackage{enumitem}
\usepackage{mathtools}
\usepackage{siunitx}

\usepackage{arydshln}
\definecolor{linkColor}{RGB}{6,125,233}
\definecolor{green}{rgb}{0.0, 0.65, 0.31}
\definecolor{bleudefrance}{rgb}{0.19, 0.55, 0.91}
\definecolor{ceruleanblue}{rgb}{0.16, 0.32, 0.75}
\definecolor{grey}{HTML}{969696}
\definecolor{violet}{HTML}{756bb1}
\definecolor{dgrey}{HTML}{01665e}
\definecolor{lgrey}{HTML}{5ab4ac}
\definecolor{dgreen}{HTML}{005a32}
\definecolor{purple}{HTML}{ae017e}

\definecolor{editCol}{HTML}{000000}
\definecolor{maskCol}{HTML}{c51b7d}
\definecolor{lrColor}{HTML}{8856a7}
\definecolor{trColor}{HTML}{d01c8b}
\definecolor{ctColor}{HTML}{4dac26}
\definecolor{brickred}{HTML}{f03b20}
\definecolor{improveCol}{HTML}{253494}
\definecolor{worsenCol}{HTML}{d7191c}
\definecolor{DarkBlue}{HTML}{00008B}
\definecolor{mscolor}{HTML}{01665e}
\definecolor{nmscolor}{HTML}{bf812d}
\definecolor{lgreen}{HTML}{ccece6}
\definecolor{dolive}{HTML}{308014}

\colorlet{tablerowcolor4}{gray!50} 

\newcommand*{\textlabel}[2]{%
  \edef\@currentlabel{#1}
  \phantomsection
  #1\label{#2}
}

\colorlet{tableheadcolor}{gray!25} 
\colorlet{tablerowcolor}{gray!10} 
\colorlet{tablerowcolor2}{gray!45} 
\colorlet{tablerowcolor3}{gray!25} 

\newcommand{\rowcollight}{\rowcolor{tablerowcolor3}} %

\newcolumntype{a}{>{\columncolor{tablerowcolor}}r}

\definecolor{aicolor}{HTML}{000000}
\definecolor{occolor}{HTML}{000000}

\def\aibarr#1{
  {\color{aicolor}\rule{#1mm}{6pt}}
  }
  
\def\ocbarr#1{
  {\color{occolor}\rule{#1mm}{6pt}}
  }

\newcommand{\hlAI}[1]{#1}
\newcommand{\hlOC}[1]{#1}

\newif{\ifhidecomments}
  \hidecommentsfalse 
\ifhidecomments
    \newcommand{\koustuv}[1]{}
    \newcommand{\munmun}[1]{}
\else
    \newcommand{\koustuv}[1]{\textbf{\small\sffamily{\textcolor{DarkBlue}{[#1 -- Koustuv]}}}}
    \newcommand{\munmun}[1]{\textbf{\small\sffamily{\textcolor{orange}{[#1 -- Munmun]}}}}
  \fi
\newcommand{\edit}[1]{{\textcolor{editCol}{#1}}}

\newcommand{\ai}{AI}
\newcommand{\oc}{OC}

\newcommand{\cdi}{$\mathtt{CDI}$}

\colorlet{tableheadcolor}{gray!25} 

\definecolor{neutralCol}{HTML}{dd1c77}
\definecolor{neutralGreen}{HTML}{31a354}
\definecolor{NewBlue}{HTML}{1879ba}
\definecolor{bleudefrance}{rgb}{0.19, 0.55, 0.91}  
\definecolor{AfTrColor}{HTML}{0868ac}  
\definecolor{BfTrColor}{HTML}{a8ddb5}  

\definecolor{AfCtColor}{HTML}{b10026}  
\definecolor{BfCtColor}{HTML}{fd8d3c}

\graphicspath{ {figures/} }

\title{Linguistic Comparison of AI- and Human-Written Responses to Online Mental Health Queries}

\author[1]{Koustuv Saha}
\author[1]{Yoshee Jain}
\author[2]{\edit{Violeta J. Rodriguez}}
\author[3]{Munmun De Choudhury}
\affil[1]{Siebel School of Computing and Data Science, The Grainger College of Engineering, University of Illinois Urbana-Champaign, Urbana, IL, USA}
\affil[2]{\edit{Department of Psychology, College of Liberal Arts and Sciences, University of Illinois Urbana-Champaign, Champaign, IL, USA}}
\affil[3]{School of Interactive Computing, College of Computing, Georgia Institute of Technology, Atlanta, GA, USA\newline}

\affil[*]{Corresponding Author:\newline
Koustuv Saha\newline
Siebel Center for Computer Science\newline
201 N. Goodwin Ave.\newline
Urbana, IL, 61801\newline
USA\newline
ksaha2@illinois.edu}


\begin{abstract}

The ubiquity and widespread use of digital and online technologies have transformed mental health support, with online mental health communities (OMHCs) providing safe spaces for peer support. 
More recently, generative AI and large language models (LLMs) have introduced new possibilities for scalable, around-the-clock mental health assistance that could potentially augment and supplement the capabilities of OMHCs. 
Although genAI shows promise in delivering immediate and personalized responses, their effectiveness in replicating the nuanced, experience-based support of human peers remains an open question. 
In this study, we harnessed 24,114 posts and 138,758 online community (\oc{}) responses from 55 OMHCs on Reddit. 
We prompted several state of the art LLMs (GPT-4-Turbo, Llama-3, and Mistral-7B) with these posts, 
and compared their responses to human-written (\oc{}) responses based on a variety of linguistic measures across psycholinguistics and lexico-semantics. 
Our findings revealed that \ai{} responses are more verbose, readable, and analytically structured, but lack linguistic diversity and personal narratives inherent in human-human interactions. 
Through a qualitative examination, we found validation as well as complementary insights into the nature of AI responses, such as its neutrality of stance and the absence of seeking back-and-forth clarifications. 
We discuss the ethical and practical implications of integrating generative AI into OMHCs, advocating for frameworks that balance AI's scalability and timeliness with the irreplaceable authenticity, social interactiveness, and expertise of human connections that form the ethos of online support communities. 

\end{abstract}
\begin{document}

\flushbottom
\maketitle

\thispagestyle{empty}



\section*{Introduction}\label{section:intro}


Advancements in digital technologies over the recent decades have been redefining how individuals engage with mental health care and support systems~\cite{de2014mental,vaidyam2019chatbots,bucci2019digital,tal2017digital,chaszczewicz2024multi}. 
Traditional therapy and peer support avenues have expanded into online spaces, offering new modes of interaction and assistance. 
Online mental health communities (OMHCs) exemplify these developments, providing safe, anonymous platforms where users can openly express their thoughts, seek advice, and connect with others experiencing similar challenges~\cite{saha2020omhc,de2014mental,sharma2018mental}.
These communities thrive on mutual support, where the collective lived experience of participants helps foster understanding and empathy. The success of OMHCs lies in their ability to create environments conducive to open self-disclosure, promising to reduce the stigma often associated with discussing mental health concerns in offline settings~\cite{sharma2018mental,de2014mental,sharma2020computational,saha2022social}. 

Recent technological developments, particularly in generative artificial intelligence (genAI), have introduced new opportunities for mental health support through sophisticated conversational agents and large language models (LLMs), above and beyond OMHCs. People are not only appropriating OMHCs for mental health help seeking, but also exploring the use of LLMs as chatbots during times of distress~\cite{sweeney2021can,sharma2024facilitating,ovsyannikova2025third,stade2024large}. Research has subsequently sought to understand the potential benefits of LLMs---GPT-4 has demonstrated the ability to mimic human-like conversation and adapt responses to users' needs~\cite{cuadra2024illusion,welivita2024chatgpt,sorin2024large}. 
Peer support is a key therapeutic approach to tackling mental health concerns~\cite{sharma2018mental,yang2019channel,kummervold2002social}, and LLMs promise to offer that kind of support around the clock in a scalable fashion extending the reach of human-based peer support available in OMHCs. This, in turn, can open doors to significantly extending and scaling mental health services, especially given the paucity of trained mental health professionals in the US~\cite{song2024typing}. 
Finally, research has shown that being able to confide in sensitive or challenging life experiences to a trusted peer can alleviate feelings of distress~\cite{davidson2012peer,shalaby2020peer,andalibi2017sensitive,sharma2018mental}, and LLMs can potentially serve as those ``trusted peers'' in OMHCs that can provide non-judgmental help and advice to people with mental health struggles. In doing so, LLMs carry the potential ability to augment and supplement support received from human peers in OMHCs.

Although these emerging AI technologies show immediate promise, LLMs as mental health support tools are yet to be thoroughly assessed against the organic, nuanced responses generated within human interactions in online communities~\cite{de2023benefits,cabrera2023improving,yoo2025ai}. 
This paper aims to bridge this gap by analyzing the lexico-semantics and overall effectiveness of 24,114 \ai{} responses generated by each of three state-of-the-art LLMs---GPT-4-Turbo, Llama-3.1, and Mistral-7B---
in comparison to those of human-written content on 55 OMHCs on Reddit.
\edit{Our study identifies previously under-explored linguistic phenomena distinguishing AI and human peer mental health support, including greater reliance on generic framing, reduced contextual grounding, and constrained expressions of emotionally attuned validation in AI responses.} Our goal is to examine how these responses compare in linguistic structure, emotional and informational support, adaptability, and potential limitations,
\edit{thereby providing one of the first population-level characterizations of how LLM-generated mental health support diverges from human-written peer mental health support at scale.}
\edit{Together, this work represents one of the first large-scale, cross-model linguistic comparisons of AI versus human mental health support in naturally occurring online contexts.}

This study offers many implications. While AI may offer scalable and empathetic support, its lack of linguistic diversity and creativity could limit its effectiveness in long-term or nuanced therapeutic contexts. The findings also emphasize 
a dual role for AI in supplementing peer support---providing consistent assistance while recognizing and addressing its gaps in personalization and context-driven empathy. This research contributes to the broader discourse on the ethical and practical integration of AI in mental health care. It underscores the necessity of developing frameworks that harness the benefits of AI's capabilities while maintaining the irreplaceable human touch of peer and professional support.

\section*{Results}\label{sec:results}

We collected 24,114 posts and 138,758 human-written (\oc{}) responses from 55 mental health-related subreddits~\cite{sharma2018mental}. To generate AI responses, we queried these posts using state-of-the-art LLMs---GPT-4-Turbo, Llama-3.1, and Mistral-7B. 
We then conducted a suite of comprehensive psycholinguistic and lexico-semantic analyses comparing \ai{} and \oc{} responses. For ease of exposition, the majority of our results focus on comparisons with GPT-4-Turbo as the representative \ai{} responses. Finally, we performed robustness analyses by extending our comparisons to Llama-3.1 and Mistral-7B.

\subsection*{Psycholinguistic Analysis}

\autoref{tab:liwc} summarizes the occurrences of psycholinguistic attributes in the \ai{} and \oc{} responses.
Although several of the comparisons are significant as per $t$-tests, we primarily focus on examining the differences which show moderate to large effect size (Cohen's $d$>0.20)~\cite{cohen2013statistical}.

\begin{itemize}
\item \textit{Affect.} \ai{} responses contained greater sadness (by 74\%) but much lower anger (by 90\%) than \oc{} responses.
This suggests that while AI responses may align with the emotional depth of distressing discussions, they might adopt a more neutral or supportive tone, minimizing confrontational or angry expressions that may present in \oc{} interactions. This is also likely associated with the heavy degree of moderation and red-teaming the LLMs have undergone to prevent too negative (or abusive) responses. 

\item \textit{Cognition and Perception.} 
\ai{} responses showed greater occurrences of \textit{differentiation} (by 68\%) and \textit{feel} (by 49\%) than \oc{} responses. 
In contrast, \ai{} responses show lower occurrences of \textit{causation} (by 47\%), \textit{certainty} (by 58\%), and \textit{see} (by 67\%). 
This plausibly indicates that \ai{} responses focus more on acknowledging emotions rather than asserting definitive explanations.


\item \textit{Social and Personal Concerns.} This category contains several content words relating to social and personal concerns. \ai{} and \oc{} responses show similar occurrences of keywords in several categories with low Cohen's $d$.
Interestingly, \ai{} responses showed lower use of \textit{friend} (-58\%), \textit{female} (-58\%), and \textit{male} (-71\%) keywords than \oc{} responses. 
These categories are associated with sharing of personal and social relationships, and in online communities, members often share their personal narratives, which may not be the case for AI. 
Again, \ai{} responses showed lower relativity-related attributes, including \textit{motion} (-50\%) and \textit{time} (-60\%) than \oc{} responses, which is also likely associated with sharing people's past experiences. 
In contrast, \ai{} responses show a significantly higher occurrence of \textit{affiliation} (25\%)  and \textit{power} (192.6\%) than \oc{} responses. This could be associated with \ai{}'s tendency to provide more structured guidance, rather than personal anecdotes.

\item \textit{Biological Concerns.} Under biological concerns, \ai{} responses showed significantly higher occurrence of \textit{health} (by 207\%), but lower occurrences of \textit{body} (by 26\%) and \textit{sexual} (by 30\%) related keywords than \oc{} responses.
This suggests that AI responses may tend to focus on health-related advice or information rather than engaging in personal or detailed interactions about physical bodies or sexual topics. In contrast, \oc{} responses may include more personal experiences, concerns, and narratives related to bodily functions and sexual health, which AI responses may avoid.

\item \textit{Function Words.} Function words are known to be associated with understanding linguistic style and psychology of expression of individuals~\cite{chung2007psychological}.
We find that \ai{} responses showed significantly greater use of articles (by 19\%), prepositions (by 14\%), auxiliary verbs (by 26\%), and conjunction (by 18\%)---all indicative of a more articulate style of writing in \ai{} responses.
In contrast, the \ai{} responses showed lower use of negation (-43\%), number (-68\%), and quantifier (-50\%). 
This suggests that \ai{} responses may avoid overly absolute or uncertain statements.

\item \textit{Interpersonal Focus (Pronouns).}
Pronoun use is associated with narrative style and focus of attention in interpersonal conversations~\cite{pennebaker2003psychological}. 
We find that \ai{} responses show a significantly lower use of first person singular (-70\%) and first person plural (-71\%) pronouns than \oc{} responses. 
This indicates the lack of personal narration style and self-attentional focus in the \ai{} responses. 
In contrast, in \oc{}s, individuals often respond by sharing their personal experiences. 
Further, the greater use of first person plural (\textit{we}, \textit{us}, etc.) reveals the sharing of experiences and solidarity as a collective identity in online communities~\cite{saha2017stress,cohn2004linguistic}. 
Again, the \ai{} responses show greater use of second-person pronouns (39\%) and impersonal pronouns (29\%) than \oc{} responses---which could associate with how \ai{} is more likely to provide structured information and advice.
Further, pronoun usage is also associated with social hierarchy---those in higher social positions are more likely to use second-person pronouns~\cite{kacewicz2014pronoun}.
Therefore, the observations on pronoun usage could be perceived as the AI assuming a higher social position than the one who asks the question.

\item \textit{Temporal References.} Temporal references in language can indicate how individuals frame their thoughts—whether they reflect on the past, anticipate the future, or focus on the present moment. We find that \ai{} responses showed lower use for \textit{past} (-62\%) and \textit{future} (-39\%) focus, but higher use of \textit{present} (15\%) focus. This suggests that \ai{} responses are less likely to engage in retrospective storytelling or speculation about future outcomes, which are common in \oc{} discussions. Instead, AI tended to provide immediate, present-focused guidance.

\item \textit{Informal Language.} Finally, we find \ai{} responses exhibited significantly lower use of the informal language across \textit{swear} (by 99\%), \textit{netspeak} (by 97\%), nonfluent (by 80\%), and \textit{filler} (by 100\%) than \oc{} responses. 
This is plausibly linked to the aspect that LLMs are trained and fine-tuned to generate formal language while avoiding inappropriate or overly casual language. 
\end{itemize}


\subsection*{Lexico-semantic Analysis}

\begin{figure}[t]
\centering
    \includegraphics[width=\columnwidth]{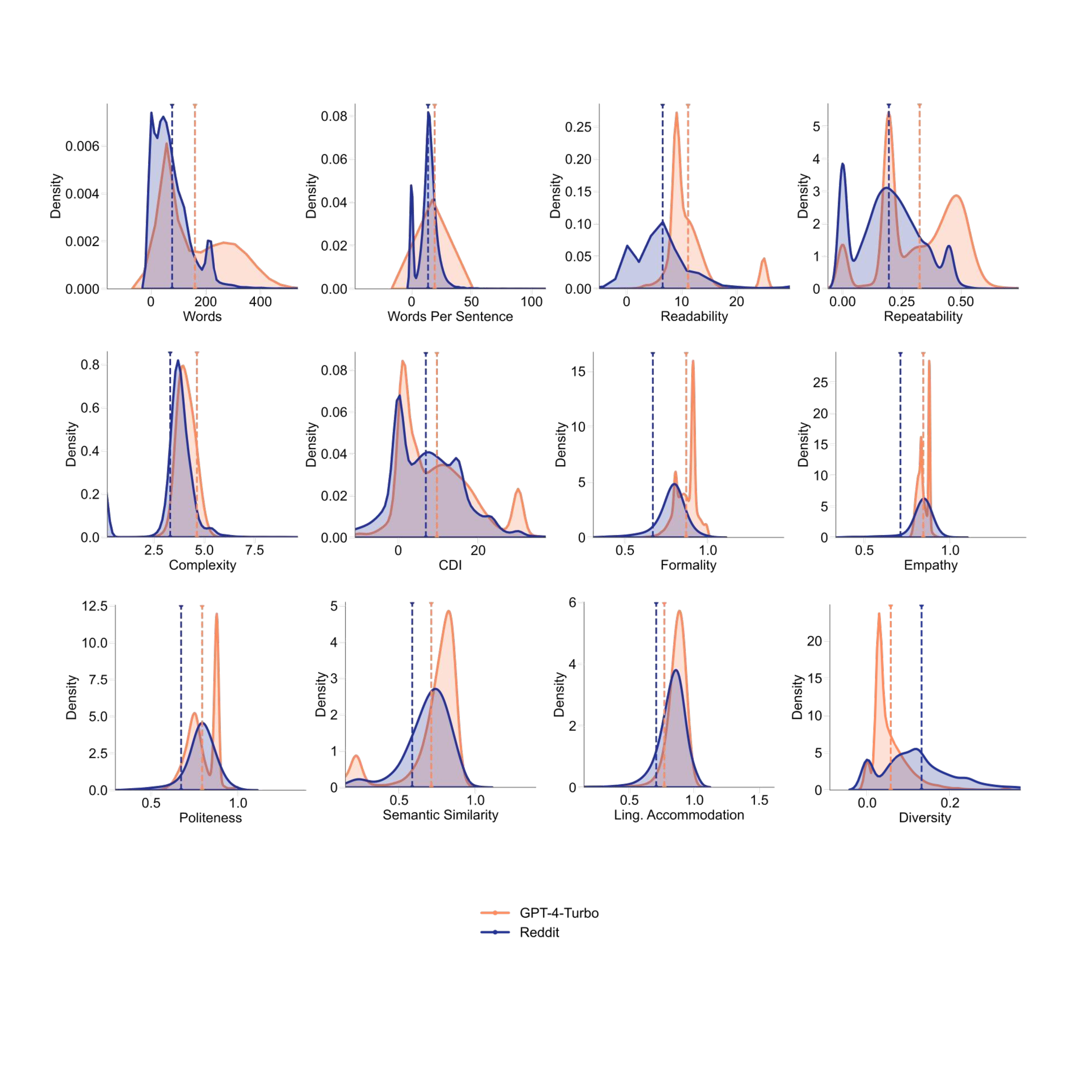}
    \caption{Comparison of the distribution of lexico-semantic measures between \textcolor{aicolor}{\textbf{GPT-4-Turbo} (\ai)} and \textcolor{occolor}{\textbf{Reddit} (\oc{})} responses. The dotted lines show respective means.}
    \label{fig:lexPDFs}
\end{figure}

We examined the lexico-semantics of language differences between the \ai{} and \oc{} responses---\autoref{tab:lexicosemantics} summarizes the results and \autoref{fig:lexPDFs} show the distributions. We describe our findings below:



Prior research highlights the critical role of \textbf{linguistic structure} in effective psychotherapy, as it shapes the depth, clarity, and impact of therapeutic communication~\cite{norcross2018psychotherapy}. 
Linguistic structure can influence both comprehension and engagement in supportive interactions. 
We examined four dimensions of linguistic structure, as described below:

\begin{itemize}
\item \textit{Verbosity} 
The depth and thoroughness of a response play a crucial role in its effectiveness in providing social support~\cite{saha2020causal}. Prior research has emphasized that both the quality and quantity of information contribute to supportive communication, with verbosity correlating to the level of detail and elaboration~\cite{glass1992quality}. 
We found statistically significant differences between \ai{} and \oc{} responses (\autoref{tab:lexicosemantics}), with \ai{} responses exhibiting greater verbosity at both the \textbf{response-level} (Cohen's $d$=0.63) and the \textbf{sentence-level} (Cohen's $d$=0.17), indicating a more detailed and lengthier style of communication in \ai{} responses.


\item \textit{Readability} \edit{refers to surface-level textual complexity commonly expressed as an estimated grade level based on structural features such as sentence length and word length~\cite{wang2013assessing,mcinnes2011readability}. We used the Coleman–Liau Index (CLI) to operationalize this measure as an indicator of linguistic complexity.}
We found that \ai{} responses showed 70\% higher readability than \oc{} responses (Cohen's $d$ = 0.71).
While this suggests that \ai{} responses (mean=11.19) exhibit higher readability, these scores correspond, \edit{under conventional interpretations of CLI,} to approximately 11 years of formal education required to engage with the text structure. In contrast, \oc{} responses have a much lower readability score (mean = 6.90), corresponding to approximately 7 years of education. 
We found that \ai{} responses showed 70\% higher readability than \oc{} responses (Cohen's $d$ = 0.71).  
\edit{Under conventional interpretation of CLI,} this suggests that \ai{} responses (mean = 11.19) exhibit a \textit{higher writing quality}, \ai{} responses also require approximately 11 years of education for comprehension. In contrast, \oc{} responses have a much lower readability score (mean = 6.90), corresponding to about 7 years of education for comprehension.

\item \textit{Repeatability} refers to the degree of repetition of words~\cite{ernala2017linguistic,saha2018social}. 
\ai{} responses showed 67\% higher repeatability than \oc{} responses (Cohen's $d$ = 0.88). While repetition can sometimes reinforce key points, in this case, it suggests a potential decrease in writing quality, especially when combined with longer or more verbose responses. This higher repeatability may reflect a redundancy in AI responses, which could impact the conciseness and clarity of communication.

\item \textit{Complexity} captures the sophistication words, based on the average length of words per sentence. \ai{} showed 40\% higher complexity than \oc{} responses (Cohen's $d$ = 0.75). The higher complexity in \ai{} responses indicates a more intricate and detailed use of language. However, the added complexity could also lead to more convoluted sentence structures that may not always be aligned for clear and simple conversational communication.

\end{itemize}

Then, we analyzed the \textbf{linguistic style of expression}, another crucial factor in the effectiveness of psychotherapy and social support~\cite{norcross2018psychotherapy,saha2020causal,cutrona1986social}.
Linguistic style influences the tone and interactional dynamics of language. We examined the differences in four dimensions of linguistic style, as described below:


\begin{itemize}
\item \textit{Categorical-Dynamic Index (CDI)} differentiates categorical (analytical, structured) and dynamic (fluid, narrative-driven) language styles~\cite{pennebaker2014small}.
We found that \ai{} responses show a 60\% higher CDI than \oc{} responses (Cohen's $d$=0.29). 
This indicates that \ai{} used a much more analytical writing style, whereas Reddit members used a personal narrative style---aligning with our observations in the psycholinguistic examination.

\item \textit{Formality} captures adherence to grammatical conventions and structured syntax, indicating the degree of professional or casual expressions~\cite{levin1991frequencies,labov1971study}.
We found that \ai{} responses showed a 30\% higher formality score than \oc{} responses with statistical significance and large effect size (Cohen's $d$=0.97).

\item \textit{Empathy} reflects the extent to which language conveys emotional understanding, validation, and engagement~\cite{herlin2016dimensions,sharma2020computational}.
Our analyses revealed that \ai{} responses demonstrated 19\% higher empathy than \oc{} responses, with a moderate effect size (Cohen's $d$ = 0.63). This suggests that \ai{} responses are more likely to incorporate linguistic cues that convey an empathetic tone. These findings align with recent research indicating that large language models (LLMs) are becoming increasingly adept at simulating empathy, creating interactions that make users feel seen and heard~\cite{inzlicht2023praise,kidder2024empathy,welivita2024chatgpt}.

\item \textit{Politeness} includes stylistic components to show respect, social harmony, conflict-avoiding, and considerate tone~\cite{brown1987politeness}. 
\ai{} responses exhibited 18\% higher politeness than \oc{} responses, with a moderate effect size (Cohen's $d$ = 0.57). This suggests that \ai{} generates more courteous and polite responses, potentially enhancing the perceived supportiveness of interactions.

\end{itemize}

We also measured \textbf{linguistic adaptability}. In interpersonal interactions, people tend to adapt to the language and expressions of each other~\cite{goffman1978presentation}. 
A body of psychotherapeutic and psycholinguistic research reveals how linguistically adaptable and accommodating responses are more effective in support than templated or generic responses~\cite{althoff2016large,de2017language,saha2020causal}.
Even for human-AI interactions, prior work noted that when an AI responds with more adaptable language to the user, the AI's perceived anthropomorphism, intelligence, and likeability are higher~\cite{wang2021mutual}. 
We examined three dimensions of linguistic adaptability as described below:


\begin{itemize}
\item \textit{Semantic Similarity} measures how semantically coherent a response is to its query. 
we found that AI responses exhibited a 21.49\% higher semantic similarity compared to OC responses with statistical significance (Cohen's $d$ = 0.52). This suggests that AI responses are more closely aligned with the specific content and intent of the query, demonstrating a greater ability to generate contextually relevant replies. 


\item \textit{Linguistic Style Accommodation} measures how well a response aligns with the query’s linguistic style, focusing on similarities of non-content words~\cite{danescu2011mark}. We found that \ai{} responses exhibited 9\% higher linguistic style accommodation than \oc{} responses (Cohen's $d$ = 0.76). This indicates that \ai{} responses can linguistically accommodate with their queries, potentially enhancing their effectiveness in facilitating supportive and engaging interactions, similar to online communities~\cite{danescu2011mark,sharma2018mental}.

\item \textit{Diversity/Creativity} refers to the uniqueness and variation of a response compared to others. Greater diversity indicates more variation in language use. We measured diversity using cosine distances from the centroid of the response set, based on word embeddings. We found that \ai{} responses had 57\% lower cosine distance from the centroid compared to \oc{} responses, with statistically significant differences (Cohen's $d$=-1.21, $t$=-90.92, $p$<0.001). This observed difference in linguistic diversity suggests that, while \ai{} may generate relevant and coherent responses, it tends to reuse the content across several responses, potentially indicating a lack of creativity in addressing individual concerns. On the other hand, online community members are likely to provide ``out-of-the-box'' suggestions based on their lived experiences. 

\end{itemize}



Finally, we measured \textbf{social support}. Social support plays a crucial role in mitigating psychological distress, acting as a protective buffer against mental health challenges~\cite{kummervold2002social,cohen1985stress}.
Online support-seeking has proven effective in reducing depression and enhancing self-efficacy and quality of life~\cite{rains2009meta}.
Again, in the context of suicide, social support within Reddit communities may lower the risk of future suicidal ideation~\cite{de2017language}.
According to the Social Support Behavioral Code~\cite{cutrona1992controllability}, two key forms of support, which have also received significant empirical and theoretical attention, are emotional support (ES) and informational support (IS).
Both of these forms of support prevalent and effective in online interactions~\cite{sharma2018mental,andalibi2018social}.

We found that \ai{} responses showed significantly higher social support---62\% higher ES and 20\% higher IS---than \oc{} responses.  
This suggests that \ai{} can generate responses that appear more supportive. However, unlike \ai{}, online community members often engage in follow-up discussions or general commentary, which may not always directly convey support.

\subsection*{Robustness of Findings and Additional Insights}


To ensure the robustness of our findings and gain further insights, we conducted two additional analyses: 1) a qualitative evaluation, \edit{2) an expert clinical evaluation of factual accuracy and safety in AI responses}, and 3) a comparison of responses from state-of-the-art LLMs.

First, we qualitatively analyzed a sample of 50 posts and their corresponding \oc{} and \ai{} responses in our dataset. 
We noted a large number of posts in which individuals struggling with mental health concerns---or their caregivers---sought informational support. We describe the key themes below:


\begin{itemize}
\item \textit{\ai{} responses lack personal narratives, unlike \oc{} responses.}
Although most of the posts sought advice, some sought connections with others who had similar lived experiences or specifically valued advice rooted in shared experiences. In responses, while \oc{} members often shared personal stories, either to alleviate their own emotional burdens or to contribute to the greater good, \ai{} provided results that were a rich source of information. Across these posts, we identified key themes highlighting similarities and differences between responses from \ai{} and \oc{} members.


\item \textit{\ai{}'s responses are structured, whereas \oc{} responses involve more conversational engagement.}
We found that \ai{} responses were formal and well-structured. 
In most of the responses, \ai{} first acknowledged the challenges faced by the user and provided both informational and emotional support.  
In contrast, the responses from \oc{} members followed a conversational style---with bidirectional communication, where the other members asked clarification and follow-up questions to provide more informed responses. Additionally, the original poster would often express gratitude and acknowledgment after receiving the support. 

\item \textit{\ai{} responses consist of standardized guidance, whereas community responses are based on personalized experiences.}
Along the lines of the above, we noted that \ai{} is likely to provide a standardized set of guidance across multiple posts, such as asking to ``consult a healthcare professional or a therapist'' or ``joining support groups. 
This echoes the findings from our quantitative analyses on the lack of diversity across responses.
In contrast, online community members often elaborated on their experience and implicitly provided advice by detailing their journey with the struggles---leading to higher diversity across responses based on distinct experiences of individuals.
Members also acknowledged the struggles of the original poster and wished well for them. 
Overall, \ai{} responses typically presented suggestions in bullet points, often reiterating similar phrases and words, whereas \oc{} comments offered more nuanced insights, often elaborating on a single suggestion through personal experience. 
This contributed to greater verbosity and repeatability in \ai{} responses, as also observed in our quantitative analyses.

\item \textit{\ai{} tends to show a neutrality in stance.}
If the author wanted to learn more about a certain product or drug, \ai{} provided both positive impact and negative side-effects, whereas the responses on OC were mostly one-sided depending on the commenter's experience with the product. If the author asked for ``experiences,'' then the responses from \ai{} included an acknowledgment that ``As an AI, I don't have personal experiences,'' and then it provided positive and negative side-effects based on its training data. In contrast, the \oc{} members included a stance that was complemented by an explanation for their stance. For example, in a post asking about ``experiences with Lexapro'', while one user responded ``Lexapro did not work for them,'' another responded ``it worked for me because it helped stop my hot flashes and tingles that I would get at heated moments''.

\item \textit{\ai{} responses have boundaries in providing experience-based support.}
In OMHCs, individuals often seek advice based on others' personal experiences with similar symptoms or treatments, such as ``Has anyone taken Zoloft? Any advice would help.'' 
While \ai{} generated a response based on summarizing reviews from its training data, it did not directly address the query, stating, ``I'm an AI and can provide general advice based on available information, but everyone's experience with medication may vary [..].''
Also, in posts where authors provided detailed accounts of their challenges and sought general advice on how to navigate life while managing their mental health struggles, \ai{} was unable to provide any suggestions. It responded with ``I'm unable to provide the help that you need'' and recommended that the author talk to a healthcare professional or a family member.



\item \textit{Despite the guardrails, \ai{} can hallucinate.}
We noted a key strength that \ai{} can often accurately recognize the abbreviations used in posts related to mental health disorders (e.g., BPD for Borderline Personality Disorder). 
In addition, we noted that the \ai{} responses are often curated with guardrails to caution the user, such as, 
``it's important to remember that I'm an AI and not a professional'' and to ``discuss any plans with a healthcare provider.'' 
Adding such warning statements ensures that users only use the advice from \ai{} to complement other resources combating problems with misleading and inaccurate responses. 
These patterns plausibly stem from the extensive moderation and red-teaming that state-of-the-art LLMs undergo before deployment.
However, we also noted examples of hallucinations in \ai{} responses. For example, one user was looking for suggestions to get their habit of ``digging out ingrown hairs from their own legs'' in control, where \ai{} first responded with ``congratulations on getting your face skin picking under control.'' Here, face skin picking was nowhere mentioned in the original post, therefore, such responses could be inaccurate and problematic.

\end{itemize}

{\color{editCol}
Next, to assess the factual correctness and potential harmfulness of AI-generated mental health responses, we conducted an expert evaluation by our clinical psychologist coauthor. A randomly sampled subset of 100 AI-generated responses was reviewed across four dimensions: \textbf{factual accuracy}, \textbf{potential harmfulness}, \textbf{emotional attunement}, and \textbf{contextual responsiveness} to users’ expressed mental health concerns. All of these dimensions were rated on a scale of 1--5, with higher values indicating a stronger presence of the rated quality. \autoref{tab:expert_ratings} reports a summary of the expert ratings.

\begin{itemize}
\item Overall, the expert evaluation found that AI-generated responses demonstrated consistently high \textbf{factual accuracy} (mean=5.00), with no identified instances of clinically incorrect or unsafe information. The responses primarily relied on general psychoeducational content rather than individualized clinical recommendations, indicating strong informational reliability but limited contextual specificity.

\item With respect to \textbf{potential harmfulness} (mean=1.14), no direct harmful advice or unsafe recommendations were observed. However, several indirect risk factors were identified, including limited acknowledgment of emotional distress in posts describing anxiety, intrusive thoughts, or interpersonal conflict; reliance on emotionally nonspecific reassurance where more tailored validation may have been beneficial; and the absence of explicitly risk-sensitive or safety-oriented language in posts that could carry implications for well-being. While these limitations do not constitute direct harm, they may introduce vulnerabilities if AI-generated guidance is interpreted without professional supervision or taken out of the appropriate clinical context.

\item The weakest ratings emerged in the dimensions of \textbf{emotional attunement} (mean=1.82) and \textbf{contextual responsiveness} (mean=2.64). \ai{} responses frequently relied on formulaic supportive phrasing or distancing language (e.g., references to being an \ai{} system) and often failed to integrate the emotional nuance or situational details present in users' narratives. Consequently, responses tended to be accurate and polite but insufficiently personalized or emotionally aligned to meet the standards of high-quality mental health support.
\end{itemize}

Therefore, this expert review indicates that while \ai{}-generated responses appear factually reliable and broadly safe, their limitations in emotional depth and contextual engagement underscore the importance of continued human clinical involvement when evaluating and deploying AI-based mental health support tools.}

Finally, having conducted our analyses with GPT-4-Turbo, we also experimented with other general-purpose LLMs, particularly Llama-3.1 and Mistral-7B. 
These three LLMs cover a spectrum of models that vary in architecture, training dataset, and optimization methods.
We summarize the psycholinguistic and lexico-semantic comparison of these models in~\autoref{tab:liwc_llms} and~\autoref{tab:lexicosemantics_llms} respectively,
including paired $t$ tests compared with \oc{} responses, and Kruskal-Wallis $H$ test across all four modalities (\oc{}, GPT, Llama, and Mistral). We notice that the trends in comparisons (by $t$-test) are very similar for all three LLMs.

\section*{Discussion}
While AI chatbots have shown initial promise, their effectiveness as mental health support tools remains largely untested against the organic and nuanced human interactions that develop in online mental health communities (OMHCs). 
This study aimed to assess how AI-generated responses compare to human-written responses in OMHCs in terms of linguistic features spanning psycholinguistics, linguistic structure, style, adaptability to query, and social support. 
We conducted our study using 24,114 posts collected from 55 OMHCs on Reddit. 
We used these posts as queries to state-of-the-art AI chatbots such as GPT-4-Turbo, Llama-3.1, and Mistral-7B, and compared the AI responses to 138,758 human-written responses in these OMHCs. 
Our analysis revealed that \ai{} responses were more verbose, readable, and complex, indicating they might be somewhat harder to comprehend than human-written responses. 
\ai{} responses tended to be more formal and structured, demonstrating higher levels of empathy and politeness. 
Notably, AI responses exhibited a predominantly analytical linguistic style, marked by greater use of articles, prepositions, and auxiliary verbs. 
In contrast, human responses followed a more narrative-driven approach, incorporating personal disclosures and solidarity expressions. 
Additionally, \ai{} responses scored higher in semantic similarity and linguistic style accommodation, indicating a greater ability to tailor language to specific queries. However, despite their linguistic richness and supportive tone, AI-generated replies lacked diversity and creative, highlighting challenges in replicating the spontaneous, experience-based advice commonly found in online mental health communities.
By conducting a deeper qualitative analysis, we found support for our quantitative linguistic analyses, as well as uncovered a number of themes of language differences in \ai{} and \oc{} responses. 
For instance, \ai{} responses tend to adopt a neutral stance of highlighting both the positives and negatives of a specific approach, whereas \oc{} responses tend to take sides. Further, people often look for prior experiences with a particular therapy or medication in OMHCs, but \ai{} lacks such experiences and tends to recommend expert/clinical care to such queries.

This study presents a computational approach based on natural language analysis to systematically evaluate the language used in AI-generated responses to mental health-related inquiries. 
These methods and the insights can guide the development of AI-assisted response writing in OMHCs, that offer personalized, empathetic, and timely interventions, ensuring they effectively meet the emotional and informational needs of individuals seeking help and advice in these spaces. Additionally, this computational framework can help identify patterns in language use, shedding light on the types of practical assistance individuals seek in OMHCs when navigating mental health challenges. 
We list these implications below:

\begin{itemize}
\item \textit{It is essential to prioritize empathy, reliability, and transparency in AI.} Our findings suggested that AI---in its current form---is more equipped to provide immediate assistance in the form of structured guidance, reinforcing its role as an informational and solution-oriented tool rather than one that shares personal experiences or future aspirations.
This aligns with prior findings on LLMs' abilities to provide empathetic responses~\cite{cuadra2024illusion,sorin2024large,sharma2024facilitating}.
These findings underscore the need to design AI tools that not only provide relevant information but also foster empathy. Unlike online community members who draw from personal experiences to offer emotional support, LLMs cannot replicate this depth of connection---as evident from our psycholinguistic and lexico-semantic analysis.
The absence of a personal narrative and a sense of belonging in AI may lead to perceptions of ``artificial'' support, potentially diminishing its effectiveness. 
This opens up discussions on whether LLM-based tools should serve primarily as informational support agents rather than as emotional support providers. In fact, ideally, end users should have the option to prioritize or disable features based on their needs. Transparency is also crucial---without clear disclosure, users with limited AI/digital literacy may mistake LLM responses to be from humans, leading to ethical concerns. A notable example is the ethical backlash against Koko---a mental health chatbot---whose users felt misled upon realizing they were interacting with AI rather than human counselors~\cite{koko_gizmodo,koko_nbc}.

\edit{
To further elaborate, building on established distinctions between \textit{cognitive empathy}, which refers to accurately recognizing and understanding another person's emotional state, and \textit{emotional empathy}, which involves resonating with and expressing affective alignment with that state~\cite{davis1983measuring,decety2004functional,batson2009these}, our findings suggest that AI-generated responses primarily demonstrate cognitive empathy through surface-level acknowledgment, summarizing user concerns, and offering general supportive language. However, AI shows clearer limitations in emotional empathy, which requires attunement to the user’s unique emotional context, personalized validation, and relational engagement. Prior work has shown that polite or supportive wording alone is not sufficient for conveying genuine care or fostering meaningful emotional connection~\cite{bickmore2005establishing,althoff2016large}. These nuances help explain why AI responses may appear supportive at the linguistic level yet feel emotionally flat or insufficiently responsive to users' lived experiences~\cite{sharma2024facilitating,yoo2025ai,dasswain2025ai}.
}

\item \textit{As AI continues to evolve, it is crucial to question what would be the role of human support will be in OMHCs in the future. }
While some users turn to these platforms for advice or resources, many seek solidarity, emotional validation, or a safe space to express thoughts they cannot share with family, caregivers, or in their offline worlds. 
Although LLM-written responses can fulfill certain informational needs, they may fall short in fostering the sense of community and human connection that many users of OMHCs value. 
The sustainability of online communities depends on active and continuous user engagement. 
If support in these platforms is increasingly AI-based and is removed from the lived experience of people, these spaces risk losing their vitality, diminishing the peer support that makes them meaningful. 
That said, the act of sharing personal experiences---whether through discussions or expressive writing---can be therapeutic, helping individuals to process emotions, gain perspective, and feel less isolated~\cite{pennebaker2007expressive,ernala2017linguistic}. 
While AI does not directly enable a similar benefit currently, it can be designed to encourage journaling, offering a private space for users to articulate their thoughts and still receive an interactive ``talking-to-someone'' experience as well as receive personalized prompts for self-care. 
However, for such interactions to be truly beneficial in OMHCs, they must be designed with user safety in mind. 
This also raises a critical question: How would the shift from human-led to AI-driven support impact mental health outcomes over time? Would it weaken the empathy and social bonds that communities provide?

\item \textit{Along the lines of the above, another pertinent question comes up: as we invest in AI-driven mental health tools, should we only prioritize efficiency over human connection, or can AI be designed to enhance, rather than replace, community-driven support?}
Although OMHCs provide platforms for individuals to share sensitive experiences and receive social support, they also present challenges such as delayed responses, exposure to online toxicity and antisocial behaviors (e.g., hateful speech, trolling, misinformation), privacy concerns, and stigmatization~\cite{cheng2015antisocial,saha2020omhc}. 
LLM based tools have the potential to mitigate some of these issues by offering immediate responses and interventions while allowing community members to provide more personalized, long-term support. 
However such approaches to integrate generative AI in response writing in OMHCs is not without concerns. 
LLMs can reproduce inappropriate stereotypes~\cite{tao2024cultural,shrawgi2024uncovering}, present clinically unvalidated  information~\cite{zhou2023synthetic}, miss cross-lingual or cross-cultural contexts~\cite{jin2024better}, and may fail to adequately understand the lived experience of an individual~\cite{chandra2024lived,yoo2025ai}. 
A hybrid human-AI model holds promise in bridging these gaps. AI can provide scalable, immediate assistance, and human 
community members can maintain the emotional depth and relational support that AI currently lacks. By carefully integrating AI, we can create a system where the technology supports and complements the core values of peer-driven mental health support.

\item \textit{Overall, as AI becomes increasingly integrated into mental health care, it is important to carefully address concerns related to biases, hallucinations, ethical dilemmas, over-reliance, and the tension between personalization and privacy~\cite{yoo2025ai,weidinger2021ethical,bender2021dangers,nori2023capabilities}.} 
For example, AI models trained on existing datasets may inadvertently reinforce cultural, racial, and gender biases, leading to inequitable support or harmful moderation practices~\cite{timmons2023call,sogancioglu2024fairness,stade2024large,kolla2024llm}. 
Additionally, AI hallucinations (as also noted in our qualitative examination)---the generation of misleading, irrelevant, or false information~\cite{zhou2023synthetic}---can be particularly dangerous in mental health contexts, potentially guiding vulnerable users toward harmful decisions. 
\edit{More broadly, our mixed findings regarding factual accuracy but limited emotional attunement underscore the need for caution in framing LLMs as replacements for human support. AI-generated responses may appear linguistically supportive while lacking deeper emotional resonance, which could create unmet expectations or false reassurance if they are used without appropriate human oversight.}

Accountability is another critical issue---when an AI offers harmful advice, who is responsible? Should the onus fall on developers, platform administrators, OMHC moderators and community members, or the AI system itself for the harm caused by automated decisions? 
Furthermore, while AI's ability to provide personalized support can greatly enhance user experiences, it also raises privacy concerns. Should AI analyze sensitive user data, often shared in OMHC postings to tailor responses, or should its adaptability be limited through strict privacy safeguards? 
It is important that users retain control over how much personal information AI systems can access and should have the option to opt out of AI-driven personalization. 
To address these concerns, robust regulation and oversight frameworks are necessary~\cite{amershi2019guidelines,ehsan2023charting,zhang2024dark}. 
\edit{Such frameworks should emphasize transparency regarding system limitations, mechanisms for continuous auditing of harmful or biased outputs, and clear accountability structures---particularly for high-stakes domains such as mental health support.}

These should ensure that AI-driven mental health support tools adhere to ethical standards, protect user privacy, and mitigate risks associated with biases and misinformation. 
\edit{In line with these concerns, responsible deployment practices, transparency about system limitations, and continued involvement of experts, clinicians, and community stakeholders are essential when designing, evaluating, and integrating AI-mediated mental health support tools.}
Striking a balance between efficiency, human connection, and ethical responsibility is crucial to ensuring that AI enhances rather than undermines mental health support in online communities.

\end{itemize}



That said, this study has \textit{limitations that suggest interesting future directions}. 
Our findings are not generalizable to the entire population or all generative AI applications, as the data is biased by self-selection---only users who chose to participate in online communities were included. 
The user base and queries for LLMs in mental health self-management may differ, highlighting the need to examine the representativeness of diverse user groups.
Additionally, the study's design, particularly how we prompted the LLMs, limits the scope of our findings. 
Responses may vary based on different and more sophisticated prompts, and we did not explore interactive, back-and-forth conversations. 
\edit{Our analyses focus on single-response interactions, consistent with the reply-based structure of online mental health communities. However, future work should investigate multi-turn conversational dynamics to better understand how empathy, personalization, trust, and safety evolve across sustained human–AI interactions.}

Furthermore, our study examined the linguistic structure, syntax, and semantics of AI responses, but did not evaluate the accuracy of AI-generated information or explore how individuals might collaborate with AI. 
\edit{To address this gap, we conducted an expert clinical evaluation of a subset of AI-generated responses. While responses were found to be factually reliable and free of direct harmful guidance, ratings showed lower emotional attunement and contextual responsiveness. These findings indicate that safety in mental health communication extends beyond informational correctness and also requires emotional sensitivity and situational alignment, and that the absence of these qualities may pose indirect risks if AI-generated guidance is used without human oversight. Future work should include user-centered studies and larger-scale clinical audits to further evaluate how emotional alignment, risk signaling, and adaptive responsiveness shape the real-world effectiveness of AI tools.}

\edit{
Our automated linguistic measures operate primarily at the sentence and response levels rather than as direct length-based counts, and responses were normalized within the analytic pipeline to reduce systematic bias from verbosity. However, response length may still influence some measured attributes, and future work can incorporate explicit length-matching or additional statistical controls to further isolate stylistic and empathetic differences post length.
}

\edit{Further, our use of automated empathy and support measures captures linguistic proxies of support rather than users’ perceived helpfulness or emotional impact. While our expert evaluation provides complementary human judgment, broader clinical and end-user assessments are needed to fully validate these metrics.}
\edit{Importantly, readability metrics such as the Coleman-Liau Index should be interpreted as indicators of surface-level linguistic complexity and informational density---derived from sentence length and word length---rather than direct measures of reader comprehension, emotional accessibility, or therapeutic effectiveness, particularly in the context of mental health support communication.}

This study motivates further investigation into the perceptions and effectiveness of such human--AI interactions, and future research could include user studies to evaluate the reliability of AI interactions and investigate user perceptions of LLM responses to health queries, thereby enhancing our understanding of how LLMs can effectively support mental health care.
An open question remains about the role of AI in mental health support---is it a friend, a peer supporter, a therapist, or merely a tool for providing information? This definition and its interpretation will likely vary among individuals, influencing the perceived effectiveness of AI in mental health care.

\section*{Methods}

\subsection*{Data and Study Design}

We sourced our data from Reddit. 
Reddit is a popular social platform consisting of online communities, called subreddits, which are dedicated to specific themes of discussions.
Prior work has noted that pseudonymity, community-driven moderation, and asynchronous peer support on Reddit help individuals overcome stigma and candidly self-disclose their sensitive mental health concerns and seek social support from community members~\cite{de2014mental,saha2020omhc,andalibi2018social}.
Based on prior work~\cite{sharma2018mental,kim2023supporters}, we identified 55 subreddits dedicated to mental health discussions (e.g., r/depression, r/anxiety, r/SuicideWatch, etc.), and collected their posts and responses between January 01, 2018 and March 31, 2024---leading to our online communities (\oc{}) dataset of 24,114 posts and 138,758 responses.
For each of these 24,114 posts, we queried OpenAI's GPT-4-Turbo model using the OpenAI API with the post body. 
In addition, we also deployed the open-source models Llama-3.1 and Mistral-7B locally and prompted them with these posts to obtain a diverse set of AI-generated responses to ensure that our findings are generalizable and applicable to a broad range of LLMs. 
\edit{Across all models, we adopted a zero-shot prompting strategy without any additional system instructions, role prompts, safety framing, or stylistic guidance. Responses were generated using default inference parameters, allowing us to attribute observed linguistic differences to model behavior rather than to prompt manipulation or tuning effects.}

\edit{We also deeply considered the ethics of our research. Although all data were drawn from publicly accessible Reddit posts, the content frequently includes highly sensitive personal disclosures. To minimize the risk of re-identification, misrepresentation, or unintended harm, we avoid publicly sharing verbatim posts and comments and report only aggregate-level results. No attempts were made to identify or contact individual users, and all analyses were conducted in accordance with ethical guidance for social media research.}


\subsection*{Analytic Approach}
To quantify the differences in linguistic measures between \ai{} and \oc{} responses, we obtained the effect size (Cohen's $d$) and evaluated statistical significance through \edit{paired} $t$-test and Kolmogorov-Smirnov (KS) test. 
\edit{All paired statistical analyses were conducted at the post level using a post-level aggregation strategy. For each post, we computed the mean value of each linguistic measure across all available \oc{} replies, leading to a single aggregate \oc{} score per post. This aggregated \oc{} score was then paired with the corresponding \ai{} response for that same post, obtaining one matched \oc{}--\ai{} observation per post for paired statistical tests. This procedure ensures independence of observations by accounting for multiple human responses within posts.}
Our metrics are inspired by a rich body of prior work in the space of online language and psychotherapy~\cite{saha2020causal,ernala2017linguistic,saha2025ai,althoff2016large,li2025emails,yim2025generative}.

A rich body of research has revealed the strong connection between language and psychosocial dynamics~\cite{pennebaker2003psychological,cohn2004linguistic,schwartz2013personality}. The effectiveness of interpersonal interactions often hinges on psycholinguistic markers, which shape self-disclosure and the support exchanged~\cite{pennebaker2003psychological}. To analyze differences in these markers between \ai{} and \oc{} responses, we leveraged the well-validated Linguistic Inquiry and Word Count (LIWC) lexicon~\cite{tausczik2010psychological}. 
LIWC categorizes language into over 60 psycholinguistic attributes, broadly categorized into eight dimensions---1) Affect, 2) Cognition and Perception, 3) Social and Personal Concerns, 4) Biological Processes, 5) Function Words, 6) Interpersonal Focus, 7) Temporal References, and 8) Informal Language.
We used the LIWC lexicon to obtain the normalized occurrences of these psycholinguistic attributes in \ai{} and \oc{} responses.

Further, we operationalized and measured several lexico-semantic dimensions, and we describe our approach below:

\begin{itemize}

\item\textit{Verbosity.}  Response length, or verbosity, has been associated with perceived quality and emotional richness in supportive interactions~\cite{glass1992quality,saha2020causal}.
To quantify verbosity, we employed two measures: (1) response-level verbosity, defined as the total number of words per response, and (2) sentence-level verbosity, measured by the average number of words per sentence. Higher values indicate more elaborate responses, whereas lower values suggest conciseness. 

\item\textit{Readability.} Readability measures the ease with which a reader can \edit{engage with the surface structure of} a text~\cite{coleman1975computer}. 
Readability has been established as an important measure within health and online health contexts, both in terms of self-expression~\cite{ernala2017linguistic,saha2020causal} as well as others' interpretation and comprehension~\cite{wang2013assessing,mcinnes2011readability}.
We adopted the Coleman-Liau Index
(CLI)~\cite{coleman1975computer,pitler2008revisiting} which assesses readability based on character and word structure within a sentence, calculated as, $\mathtt{CLI = (0.0588L-0.296S-15.8)}$, where $\mathtt{L}$ is the average number of letters per 100 words, and $\mathtt{S}$ is the average number of sentences per 100 words. 
\edit{While higher CLI values are linked to better writing quality, and are conventionally interpreted as corresponding to higher grade-level reading demands, they should not be treated as direct measures of reader comprehension.}


\item\textit{Repeatability and Complexity.} 
Repeatability and complexity are syntactic measures that capture the richness and depth of expression in communication, and are linked to one's cognitive state through planning, execution, and memory~\cite{ernala2017linguistic,saha2018social}.
\textit{Repeatability} accounts for the frequency with which words are repeated or reused in a piece of text. A higher degree of repetition is often associated with lower linguistic diversity, and reduced content quality.
\textit{Complexity} shapes the nature of communication~\cite{kolden2011congruence}, as a more linguistically complex text tends to convey greater nuance, precision, and depth in expressing ideas or information.
Drawing on prior work~\cite{ernala2017linguistic,saha2020causal,yuan2023mental}, we measured repeatability as the normalized occurrence of non-unique words and complexity as the average length of words per sentence. 

\item\textit{Categorical-Dynamic Index (CDI).} 
Language style can be conceptualized as existing on a continuum between categorical and dynamic modes~\cite{pennebaker2014small}. Categorical language reflects a more structured, analytical approach, akin to an ``amateur scientist'' style, where the focus is on logically organized, abstract concepts and detailed categorization. In contrast, dynamic language is more fluid and personal, commonly seen in socially engaged individuals who convey stories and reflect more on their immediate environment, incorporating personal narratives and emotional expressions. This spectrum of language use is captured by the Categorical-Dynamic Index (CDI), a bipolar measure that quantifies the balance between these two styles. A higher CDI score corresponds to a more categorical style, while a lower score indicates a dynamic, narrative-driven approach~\cite{pennebaker2014small}. The CDI is calculated based on the frequency of specific style-related parts of speech in a given text, with the formula:

\noindent{\small \cdi{} = (30 $+$ \textit{article $+$ preposition $-$ personal pronoun $-$ impersonal pronoun $-$ aux. verb $-$ conjunction $-$ adverb $-$ negation})}

To calculate the above, we obtained the occurrences of these parts of speech using the LIWC lexicon~\cite{tausczik2010psychological}. 

\item\textit{Formality.} 
Formality is a key sociolinguistic construct which is known to vary across cultures, contexts, and audiences~\cite{levin1991frequencies,heylighen1999formality,labov1971study}.
Linguistic formality encompasses the level of sophistication, politeness, and adherence to established linguistic conventions in written communication~\cite{larsson2020syntactic}. Formal language is marked by grammatical precision, structured syntax, and appropriate vocabulary, making it prevalent in professional settings, academic writing, official documentation, and respectful discourse. In contrast, informal language adopts a more relaxed and conversational tone, often incorporating slang, colloquialisms, and contractions, making it more common in casual or social interactions~\cite{li2025emails,heylighen1999formality}.
To measure formality, we leveraged a RoBERTa-based formality classifier from prior work~\cite{babakov2023don}. 
This classifier is built on Grammarly's Yahoo Answers Formality Corpus (GYAFC)~\cite{rao2018dear} and the Online Formality Corpus~\cite{pavlick2016empirical}, and it achieves an approximate accuracy of 91\% on its benchmark dataset. 
For any given text, this classifier outputs a score between 0 to 1---a higher score indicating a greater degree of formal language. 
We employed this classifier on our \ai{} and \oc{} response datasets and compared the formality scores. 

\item\textit{Empathy.} 
Given our specific setting, empathy is a key mechanism in providing support~\cite{herlin2016dimensions,sharma2020computational,sharma2023human}. 
Empathy refers to a cognitively complex process in which one can stand in the shoes of another person to understand their perspective, emotions, and the situations they are in~\cite{herlin2016dimensions}.
Prior work evaluated the effectiveness of artificially created empathetic conversations by a chatbot~\cite{morris2018towards} and noted the success and positive reception of empathetic responses in online interaction settings~\cite{sharma2020computational}.
We leveraged a RoBERTa-based empathy detection model, finetuned on a dataset of empathetic reactions to news stories~\cite{buechel2018modeling,tafreshi2021wassa}. 
Given a text, this model predicts the empathetic nature of a text---higher scores indicate greater expression of empathy. 

\item\textit{Politeness.} 
Politeness is essential in therapeutic conversations, as it helps foster a supportive, respectful, and empathetic environment that fosters trust and openness~\cite{brown1987politeness}.
To assess politeness levels, we leveraged a pre-trained politeness classification model~\cite{srinivasan2022tydip}, which assigned politeness scores ranging from 0 to 1 to both \ai{} and \oc{} responses.

\edit{
To assess the applicability of the classifiers of formality, empathy, and politeness in our dataset, we conducted a sanity check on a randomly sampled set of 185 \ai{} responses. We found that labels demonstrated high agreement with human judgments, with accuracy rates of 98.9\% for formality (183/185), 98.4\% for empathy (182/185), and 98.4\% for politeness (182/185). This provides additional confidence in the reliability of our automated measures within the mental-health context.}

\item\textit{Semantic Similarity.} 
Semantic similarity measures how well a response aligns with the underlying meaning and intent of the original query, reflecting the degree to which the response addresses the core content of the query in a coherent and contextually relevant manner. 
We obtained the semantic similarity between a query and a response by measuring the pairwise cosine similarity of word embedding representations of the queries and responses---where word embeddings are vector representations of words in latent lexico-semantic dimensions~\cite{mikolov2013distributed,pennington2014glove}. 
We used the 300-dimensional pre-trained word embeddings, trained on word-word co-occurrences in the Google News dataset containing about 100 billion words~\cite{mikolov2013distributed}.

\item\textit{Linguistic Style Accommodation.} 
While the above semantic similarity measure considers the content-based similarity, linguistic style accommodation focuses on non-content similarity---or how well a response accommodates the linguistic style of its query.
More linguistically accommodating responses are known to be associated with more effective online support~\cite{sharma2018mental,saha2020causal}.
We obtained the linguistic style accommodation between a query and a response by obtaining the cosine similarity of the vectors based on the normalized occurrences of linguistic style dimensions of articles, prepositions, pronouns, auxiliary verbs, conjunctions, adverbs, negations etc, as obtained by using the LIWC lexicon~\cite{pennebaker2003psychological,saha2020causal}. 

\item\textit{Diversity/Creativity.} Diverse and creative responses are known to be effective in psychotherapy and social support~\cite{althoff2016large,norcross2018psychotherapy,saha2020causal}.
We measured diversity by leveraging word-embedding representations in the 300-dimensional space~\cite{mikolov2013distributed}.
Within each of the \ai{} and \oc{} datasets, we first obtained the centroid vectors by averaging the word embeddings of the responses. 
Then, we iterated through each of the responses within the two datasets and measured the cosine distance from the corresponding centroids.
This distance measure signals how diverse (or creative) a particular response is to an average of all the responses---the greater the distance, the higher the diversity.

\item\textit{Social Support.} 
Online social support, including emotional (empathy, encouragement) and informational (guidance, advice) support, plays a vital role in reducing psychological distress and improving mental health~\cite{kummervold2002social,cohen1985stress,de2017language,sharma2018mental}.
To identify support expressions in the responses, we used an expert-appraised dataset and classifier built in prior work~\cite{sharma2018mental,saha2020causal}. 
These are binary SVM classifiers that assess the degree (high/low) of emotional (ES) and informational support (IS) in social media data. 
These classifiers were expert-appraised in prior work~\cite{sharma2018mental}, demonstrating strong performance---achieving $k$-fold validation accuracies of 0.71 for ES and 0.77 for IS~\cite{saha2020causal}. We leveraged these classifiers to label the presence of ES and IS in the responses.

\end{itemize}


\subsection*{Qualitative Analysis of Smaller Sample of Data \edit{and Expert Evaluation}}

We conducted a qualitative analysis on a randomly selected sample of 50 posts and their corresponding \oc{} and \ai{} (GPT-4-Turbo) responses from our dataset.
\edit{Using an inductive approach, two co-authors first independently performed open coding on an initial subset of 10 posts to surface emergent concepts and develop preliminary interpretive themes. The co-authors then met to conduct collaborative affinity diagramming and thematic synthesis, discussing points of convergence and divergence, reconciling interpretations, and iteratively refining candidate themes through consensus-based discussion. Following this, the second author completed coding of the remaining posts, continuing to apply and refine themes as additional patterns emerged. We then conducted a thematic analysis to integrate codes into higher-level categories, yielding a set of coherent themes characterizing systematic similarities and differences between \ai{} and \oc{} responses.} We ensured that each of the 50 posts included at least one corresponding \oc{} response to support direct comparative analysis between human and \ai{} responses.

\edit{In addition to this qualitative comparison, we conducted an \textbf{expert clinical evaluation} on a randomly sampled set of 100 \ai{}-generated responses. Our psychologist coauthor independently rated each response across four dimensions: \textbf{factual accuracy}, \textbf{potential harmfulness}, \textbf{emotional attunement}, and \textbf{contextual responsiveness} to users’ expressed mental health concerns. All dimensions were rated on a 1--5 scale, with higher values indicating a greater presence of the rated quality.}




\section*{Data Availability}
\edit{The datasets generated and/or analyzed during the current study are not publicly available due to the need to protect the privacy of users disclosing mental health concerns in online communities and to mitigate risks of ethical misuse of sensitive content. However, the data are available from the corresponding author on reasonable request, subject to appropriate ethical safeguards and data use agreements where applicable.}

\section*{Acknowledgments}
MDC was partly supported through NIH grants R01MH117172 and P50MH115838, and a grant from the American Foundation for Suicide Prevention. We are also grateful to Microsoft for supporting this research through the ``Foundation Models Academic Research'' program.
We thank Sachin Pendse for sharing the Reddit online mental health community dataset and Jiawei Zhou, Shruti Belgali, Navya Khanna, and Sachin Pendse for participating in initial conversations of the work.

\section*{Author contributions statement}
KS formulated the problem; KS and MDC designed the research; KS and MDC conceptualized and developed the analytic techniques; KS and YJ gathered and analyzed the data; KS, YJ, and VJR interpreted the results; \edit{VJR contributed clinical expertise and evaluation}; All authors drafted and edited the paper.

\section*{Competing interests statement}
The authors declare no competing interests.

\bibliography{0paper}

\begin{thebibliography}{100}
\urlstyle{rm}
\expandafter\ifx\csname url\endcsname\relax
  \def\url#1{\texttt{#1}}\fi
\expandafter\ifx\csname urlprefix\endcsname\relax\def\urlprefix{URL }\fi
\expandafter\ifx\csname doiprefix\endcsname\relax\def\doiprefix{DOI: }\fi
\providecommand{\bibinfo}[2]{#2}
\providecommand{\eprint}[2][]{\url{#2}}

\bibitem{de2014mental}
\bibinfo{author}{De~Choudhury, M.} \& \bibinfo{author}{De, S.}
\newblock \bibinfo{title}{Mental health discourse on reddit: Self-disclosure, social support, and anonymity}.
\newblock In \emph{\bibinfo{booktitle}{Proceedings of the international AAAI conference on web and social media}}, vol.~\bibinfo{volume}{8}, \bibinfo{pages}{71--80} (\bibinfo{year}{2014}).

\bibitem{vaidyam2019chatbots}
\bibinfo{author}{Vaidyam, A.~N.}, \bibinfo{author}{Wisniewski, H.}, \bibinfo{author}{Halamka, J.~D.}, \bibinfo{author}{Kashavan, M.~S.} \& \bibinfo{author}{Torous, J.~B.}
\newblock \bibinfo{journal}{\bibinfo{title}{Chatbots and conversational agents in mental health: a review of the psychiatric landscape}}.
\newblock {\emph{\JournalTitle{The Canadian Journal of Psychiatry}}} \textbf{\bibinfo{volume}{64}}, \bibinfo{pages}{456--464} (\bibinfo{year}{2019}).

\bibitem{bucci2019digital}
\bibinfo{author}{Bucci, S.}, \bibinfo{author}{Schwannauer, M.} \& \bibinfo{author}{Berry, N.}
\newblock \bibinfo{journal}{\bibinfo{title}{The digital revolution and its impact on mental health care}}.
\newblock {\emph{\JournalTitle{Psychology and Psychotherapy: Theory, Research and Practice}}} \textbf{\bibinfo{volume}{92}}, \bibinfo{pages}{277--297} (\bibinfo{year}{2019}).

\bibitem{tal2017digital}
\bibinfo{author}{Tal, A.} \& \bibinfo{author}{Torous, J.}
\newblock \bibinfo{journal}{\bibinfo{title}{The digital mental health revolution: Opportunities and risks}}.
\newblock {\emph{\JournalTitle{Psychiatric rehabilitation journal}}} \textbf{\bibinfo{volume}{40}}, \bibinfo{pages}{263--265} (\bibinfo{year}{2017}).

\bibitem{chaszczewicz2024multi}
\bibinfo{author}{Chaszczewicz, A.} \emph{et~al.}
\newblock \bibinfo{title}{Multi-level feedback generation with large language models for empowering novice peer counselors}.
\newblock In \emph{\bibinfo{booktitle}{Proceedings of the 62nd Annual Meeting of the Association for Computational Linguistics (Volume 1: Long Papers)}} (\bibinfo{year}{2024}).

\bibitem{saha2020omhc}
\bibinfo{author}{Saha, K.}, \bibinfo{author}{Ernala, S.~K.}, \bibinfo{author}{Dutta, S.}, \bibinfo{author}{Sharma, E.} \& \bibinfo{author}{De~Choudhury, M.}
\newblock \bibinfo{title}{Understanding moderation in online mental health communities}.
\newblock In \emph{\bibinfo{booktitle}{International Conference on Human-Computer Interaction}}, \bibinfo{pages}{87--107} (\bibinfo{organization}{Springer}, \bibinfo{year}{2020}).

\bibitem{sharma2018mental}
\bibinfo{author}{Sharma, E.} \& \bibinfo{author}{De~Choudhury, M.}
\newblock \bibinfo{title}{Mental health support and its relationship to linguistic accommodation in online communities}.
\newblock In \emph{\bibinfo{booktitle}{CHI}} (\bibinfo{year}{2018}).

\bibitem{sharma2020computational}
\bibinfo{author}{Sharma, A.}, \bibinfo{author}{Miner, A.}, \bibinfo{author}{Atkins, D.} \& \bibinfo{author}{Althoff, T.}
\newblock \bibinfo{title}{A computational approach to understanding empathy expressed in text-based mental health support}.
\newblock In \emph{\bibinfo{booktitle}{Proceedings of the 2020 Conference on Empirical Methods in Natural Language Processing (EMNLP)}}, \bibinfo{pages}{5263--5276} (\bibinfo{year}{2020}).

\bibitem{saha2022social}
\bibinfo{author}{Saha, K.}, \bibinfo{author}{Yousuf, A.}, \bibinfo{author}{Boyd, R.~L.}, \bibinfo{author}{Pennebaker, J.~W.} \& \bibinfo{author}{De~Choudhury, M.}
\newblock \bibinfo{journal}{\bibinfo{title}{Social media discussions predict mental health consultations on college campuses}}.
\newblock {\emph{\JournalTitle{Scientific reports}}} \textbf{\bibinfo{volume}{12}}, \bibinfo{pages}{123} (\bibinfo{year}{2022}).

\bibitem{sweeney2021can}
\bibinfo{author}{Sweeney, C.} \emph{et~al.}
\newblock \bibinfo{journal}{\bibinfo{title}{Can chatbots help support a person’s mental health? perceptions and views from mental healthcare professionals and experts}}.
\newblock {\emph{\JournalTitle{ACM Transactions on Computing for Healthcare}}} \textbf{\bibinfo{volume}{2}}, \bibinfo{pages}{1--15} (\bibinfo{year}{2021}).

\bibitem{sharma2024facilitating}
\bibinfo{author}{Sharma, A.}, \bibinfo{author}{Rushton, K.}, \bibinfo{author}{Lin, I.~W.}, \bibinfo{author}{Nguyen, T.} \& \bibinfo{author}{Althoff, T.}
\newblock \bibinfo{title}{Facilitating self-guided mental health interventions through human-language model interaction: A case study of cognitive restructuring}.
\newblock In \emph{\bibinfo{booktitle}{Proceedings of the 2024 CHI Conference on Human Factors in Computing Systems}}, \bibinfo{pages}{1--29} (\bibinfo{year}{2024}).

\bibitem{ovsyannikova2025third}
\bibinfo{author}{Ovsyannikova, D.}, \bibinfo{author}{de~Mello, V.~O.} \& \bibinfo{author}{Inzlicht, M.}
\newblock \bibinfo{journal}{\bibinfo{title}{Third-party evaluators perceive ai as more compassionate than expert humans}}.
\newblock {\emph{\JournalTitle{Communications Psychology}}} \textbf{\bibinfo{volume}{3}}, \bibinfo{pages}{4} (\bibinfo{year}{2025}).

\bibitem{stade2024large}
\bibinfo{author}{Stade, E.~C.} \emph{et~al.}
\newblock \bibinfo{journal}{\bibinfo{title}{Large language models could change the future of behavioral healthcare: a proposal for responsible development and evaluation}}.
\newblock {\emph{\JournalTitle{NPJ Mental Health Research}}} \textbf{\bibinfo{volume}{3}}, \bibinfo{pages}{12} (\bibinfo{year}{2024}).

\bibitem{cuadra2024illusion}
\bibinfo{author}{Cuadra, A.} \emph{et~al.}
\newblock \bibinfo{title}{The illusion of empathy? notes on displays of emotion in human-computer interaction}.
\newblock In \emph{\bibinfo{booktitle}{Proceedings of the 2024 CHI Conference on Human Factors in Computing Systems}}, \bibinfo{pages}{1--18} (\bibinfo{year}{2024}).

\bibitem{welivita2024chatgpt}
\bibinfo{author}{Welivita, A.} \& \bibinfo{author}{Pu, P.}
\newblock \bibinfo{journal}{\bibinfo{title}{Is chatgpt more empathetic than humans?}}
\newblock {\emph{\JournalTitle{arXiv preprint arXiv:2403.05572}}}  (\bibinfo{year}{2024}).

\bibitem{sorin2024large}
\bibinfo{author}{Sorin, V.} \emph{et~al.}
\newblock \bibinfo{journal}{\bibinfo{title}{Large language models and empathy: Systematic review}}.
\newblock {\emph{\JournalTitle{Journal of Medical Internet Research}}} \textbf{\bibinfo{volume}{26}}, \bibinfo{pages}{e52597} (\bibinfo{year}{2024}).

\bibitem{yang2019channel}
\bibinfo{author}{Yang, D.}, \bibinfo{author}{Yao, Z.}, \bibinfo{author}{Seering, J.} \& \bibinfo{author}{Kraut, R.}
\newblock \bibinfo{title}{The channel matters: Self-disclosure, reciprocity and social support in online cancer support groups}.
\newblock In \emph{\bibinfo{booktitle}{Proceedings of the 2019 chi conference on human factors in computing systems}}, \bibinfo{pages}{1--15} (\bibinfo{year}{2019}).

\bibitem{kummervold2002social}
\bibinfo{author}{Kummervold, P.~E.} \emph{et~al.}
\newblock \bibinfo{journal}{\bibinfo{title}{Social support in a wired world: use of online mental health forums in norway}}.
\newblock {\emph{\JournalTitle{Nordic journal of psychiatry}}}  (\bibinfo{year}{2002}).

\bibitem{song2024typing}
\bibinfo{author}{Song, I.}, \bibinfo{author}{Pendse, S.~R.}, \bibinfo{author}{Kumar, N.} \& \bibinfo{author}{De~Choudhury, M.}
\newblock \bibinfo{journal}{\bibinfo{title}{The typing cure: Experiences with large language model chatbots for mental health support}}.
\newblock {\emph{\JournalTitle{arXiv preprint arXiv:2401.14362}}}  (\bibinfo{year}{2024}).

\bibitem{davidson2012peer}
\bibinfo{author}{Davidson, L.}, \bibinfo{author}{Guy, K.} \emph{et~al.}
\newblock \bibinfo{journal}{\bibinfo{title}{Peer support among persons with severe mental illnesses: a review of evidence and experience}}.
\newblock {\emph{\JournalTitle{World psychiatry}}} \textbf{\bibinfo{volume}{11}}, \bibinfo{pages}{123--128} (\bibinfo{year}{2012}).

\bibitem{shalaby2020peer}
\bibinfo{author}{Shalaby, R. A.~H.} \& \bibinfo{author}{Agyapong, V.~I.}
\newblock \bibinfo{journal}{\bibinfo{title}{Peer support in mental health: literature review}}.
\newblock {\emph{\JournalTitle{JMIR mental health}}} \textbf{\bibinfo{volume}{7}}, \bibinfo{pages}{e15572} (\bibinfo{year}{2020}).

\bibitem{andalibi2017sensitive}
\bibinfo{author}{Andalibi, N.}, \bibinfo{author}{Ozturk, P.} \& \bibinfo{author}{Forte, A.}
\newblock \bibinfo{title}{Sensitive self-disclosures, responses, and social support on instagram: the case of\# depression}.
\newblock In \emph{\bibinfo{booktitle}{Proceedings of the 2017 ACM conference on computer supported cooperative work and social computing}}, \bibinfo{pages}{1485--1500} (\bibinfo{year}{2017}).

\bibitem{de2023benefits}
\bibinfo{author}{De~Choudhury, M.}, \bibinfo{author}{Pendse, S.~R.} \& \bibinfo{author}{Kumar, N.}
\newblock \bibinfo{journal}{\bibinfo{title}{Benefits and harms of large language models in digital mental health}}.
\newblock {\emph{\JournalTitle{arXiv preprint arXiv:2311.14693}}}  (\bibinfo{year}{2023}).

\bibitem{cabrera2023improving}
\bibinfo{author}{Cabrera, {\'A}.~A.}, \bibinfo{author}{Perer, A.} \& \bibinfo{author}{Hong, J.~I.}
\newblock \bibinfo{journal}{\bibinfo{title}{Improving human-ai collaboration with descriptions of ai behavior}}.
\newblock {\emph{\JournalTitle{Proceedings of the ACM on Human-Computer Interaction}}} \textbf{\bibinfo{volume}{7}}, \bibinfo{pages}{1--21} (\bibinfo{year}{2023}).

\bibitem{yoo2025ai}
\bibinfo{author}{Yoo, D.~W.}, \bibinfo{author}{Shi, J.~M.}, \bibinfo{author}{Rodriguez, V.~J.} \& \bibinfo{author}{Saha, K.}
\newblock \bibinfo{journal}{\bibinfo{title}{Ai chatbots for mental health: Values and harms from lived experiences of depression}}.
\newblock {\emph{\JournalTitle{arXiv preprint arXiv:2504.18932}}}  (\bibinfo{year}{2025}).

\bibitem{cohen2013statistical}
\bibinfo{author}{Cohen, J.}
\newblock \emph{\bibinfo{title}{Statistical power analysis for the behavioral sciences}} (\bibinfo{publisher}{Routledge}, \bibinfo{year}{2013}).

\bibitem{chung2007psychological}
\bibinfo{author}{Chung, C.} \& \bibinfo{author}{Pennebaker, J.~W.}
\newblock \bibinfo{journal}{\bibinfo{title}{The psychological functions of function words}}.
\newblock {\emph{\JournalTitle{Social communication}}} \bibinfo{pages}{343--359} (\bibinfo{year}{2007}).

\bibitem{pennebaker2003psychological}
\bibinfo{author}{Pennebaker, J.~W.}, \bibinfo{author}{Mehl, M.~R.} \& \bibinfo{author}{Niederhoffer, K.~G.}
\newblock \bibinfo{journal}{\bibinfo{title}{Psychological aspects of natural language use: Our words, our selves}}.
\newblock {\emph{\JournalTitle{Annual review of psychology}}} \textbf{\bibinfo{volume}{54}}, \bibinfo{pages}{547--577} (\bibinfo{year}{2003}).

\bibitem{saha2017stress}
\bibinfo{author}{Saha, K.} \& \bibinfo{author}{De~Choudhury, M.}
\newblock \bibinfo{journal}{\bibinfo{title}{Modeling stress with social media around incidents of gun violence on college campuses}}.
\newblock {\emph{\JournalTitle{Proceedings of the ACM on Human-Computer Interaction}}} \textbf{\bibinfo{volume}{1}}, \bibinfo{pages}{92} (\bibinfo{year}{2017}).

\bibitem{cohn2004linguistic}
\bibinfo{author}{Cohn, M.~A.}, \bibinfo{author}{Mehl, M.~R.} \& \bibinfo{author}{Pennebaker, J.~W.}
\newblock \bibinfo{journal}{\bibinfo{title}{Linguistic markers of psychological change surrounding september 11, 2001}}.
\newblock {\emph{\JournalTitle{Psychological science}}} \textbf{\bibinfo{volume}{15}}, \bibinfo{pages}{687--693} (\bibinfo{year}{2004}).

\bibitem{kacewicz2014pronoun}
\bibinfo{author}{Kacewicz, E.}, \bibinfo{author}{Pennebaker, J.~W.}, \bibinfo{author}{Davis, M.}, \bibinfo{author}{Jeon, M.} \& \bibinfo{author}{Graesser, A.~C.}
\newblock \bibinfo{journal}{\bibinfo{title}{Pronoun use reflects standings in social hierarchies}}.
\newblock {\emph{\JournalTitle{Journal of Language and Social Psychology}}} \textbf{\bibinfo{volume}{33}}, \bibinfo{pages}{125--143} (\bibinfo{year}{2014}).

\bibitem{norcross2018psychotherapy}
\bibinfo{author}{Norcross, J.~C.} \& \bibinfo{author}{Lambert, M.~J.}
\newblock \bibinfo{journal}{\bibinfo{title}{Psychotherapy relationships that work iii.}}
\newblock {\emph{\JournalTitle{Psychotherapy}}} \textbf{\bibinfo{volume}{55}}, \bibinfo{pages}{303} (\bibinfo{year}{2018}).

\bibitem{saha2020causal}
\bibinfo{author}{Saha, K.} \& \bibinfo{author}{Sharma, A.}
\newblock \bibinfo{title}{Causal factors of effective psychosocial outcomes in online mental health communities}.
\newblock In \emph{\bibinfo{booktitle}{Proceedings of the international AAAI conference on web and social media}}, vol.~\bibinfo{volume}{14}, \bibinfo{pages}{590--601} (\bibinfo{year}{2020}).

\bibitem{glass1992quality}
\bibinfo{author}{Glass, T.~A.} \& \bibinfo{author}{Maddox, G.~L.}
\newblock \bibinfo{journal}{\bibinfo{title}{The quality and quantity of social support: stroke recovery as psycho-social transition}}.
\newblock {\emph{\JournalTitle{Soc. Sci. Med.}}}  (\bibinfo{year}{1992}).

\bibitem{wang2013assessing}
\bibinfo{author}{Wang, L.-W.}, \bibinfo{author}{Miller, M.~J.}, \bibinfo{author}{Schmitt, M.~R.} \& \bibinfo{author}{Wen, F.~K.}
\newblock \bibinfo{journal}{\bibinfo{title}{Assessing readability formula differences with written health information materials: application, results, and recommendations}}.
\newblock {\emph{\JournalTitle{Research in Social and Administrative Pharmacy}}} \textbf{\bibinfo{volume}{9}}, \bibinfo{pages}{503--516} (\bibinfo{year}{2013}).

\bibitem{mcinnes2011readability}
\bibinfo{author}{Mcinnes, N.} \& \bibinfo{author}{Haglund, B.~J.}
\newblock \bibinfo{journal}{\bibinfo{title}{Readability of online health information: implications for health literacy}}.
\newblock {\emph{\JournalTitle{Informatics for health and social care}}} \textbf{\bibinfo{volume}{36}}, \bibinfo{pages}{173--189} (\bibinfo{year}{2011}).

\bibitem{ernala2017linguistic}
\bibinfo{author}{Ernala, S.~K.}, \bibinfo{author}{Rizvi, A.~F.}, \bibinfo{author}{Birnbaum, M.~L.}, \bibinfo{author}{Kane, J.~M.} \& \bibinfo{author}{De~Choudhury, M.}
\newblock \bibinfo{journal}{\bibinfo{title}{Linguistic markers indicating therapeutic outcomes of social media disclosures of schizophrenia}}.
\newblock {\emph{\JournalTitle{Proceedings of the ACM on Human-Computer Interaction}}} \textbf{\bibinfo{volume}{1}}, \bibinfo{pages}{1--27} (\bibinfo{year}{2017}).

\bibitem{saha2018social}
\bibinfo{author}{Saha, K.}, \bibinfo{author}{Weber, I.} \& \bibinfo{author}{De~Choudhury, M.}
\newblock \bibinfo{title}{A social media based examination of the effects of counseling recommendations after student deaths on college campuses}.
\newblock In \emph{\bibinfo{booktitle}{Twelfth International AAAI Conference on Web and Social Media}} (\bibinfo{year}{2018}).

\bibitem{cutrona1986social}
\bibinfo{author}{Cutrona, C.~E.} \& \bibinfo{author}{Troutman, B.~R.}
\newblock \bibinfo{journal}{\bibinfo{title}{Social support, infant temperament, and parenting self-efficacy: A mediational model of postpartum depression}}.
\newblock {\emph{\JournalTitle{Child development}}} \bibinfo{pages}{1507--1518} (\bibinfo{year}{1986}).

\bibitem{pennebaker2014small}
\bibinfo{author}{Pennebaker, J.~W.}, \bibinfo{author}{Chung, C.~K.}, \bibinfo{author}{Frazee, J.}, \bibinfo{author}{Lavergne, G.~M.} \& \bibinfo{author}{Beaver, D.~I.}
\newblock \bibinfo{journal}{\bibinfo{title}{When small words foretell academic success: The case of college admissions essays}}.
\newblock {\emph{\JournalTitle{PloS one}}} \textbf{\bibinfo{volume}{9}}, \bibinfo{pages}{e115844} (\bibinfo{year}{2014}).

\bibitem{levin1991frequencies}
\bibinfo{author}{Levin, H.} \& \bibinfo{author}{Novak, M.}
\newblock \bibinfo{journal}{\bibinfo{title}{Frequencies of latinate and germanic words in english as determinants of formality}}.
\newblock {\emph{\JournalTitle{Discourse Processes}}} \textbf{\bibinfo{volume}{14}}, \bibinfo{pages}{389--398} (\bibinfo{year}{1991}).

\bibitem{labov1971study}
\bibinfo{author}{Labov, W.} \emph{et~al.}
\newblock \bibinfo{journal}{\bibinfo{title}{The study of language in its social context}}.
\newblock {\emph{\JournalTitle{Advances in the Sociology of Language}}} \textbf{\bibinfo{volume}{1}} (\bibinfo{year}{1971}).

\bibitem{herlin2016dimensions}
\bibinfo{author}{Herlin, I.} \& \bibinfo{author}{Visap{\"a}{\"a}, L.}
\newblock \bibinfo{journal}{\bibinfo{title}{Dimensions of empathy in relation to language}}.
\newblock {\emph{\JournalTitle{Nordic Journal of linguistics}}} \textbf{\bibinfo{volume}{39}}, \bibinfo{pages}{135--157} (\bibinfo{year}{2016}).

\bibitem{inzlicht2023praise}
\bibinfo{author}{Inzlicht, M.}, \bibinfo{author}{Cameron, C.~D.}, \bibinfo{author}{D’Cruz, J.} \& \bibinfo{author}{Bloom, P.}
\newblock \bibinfo{journal}{\bibinfo{title}{In praise of empathic ai}}.
\newblock {\emph{\JournalTitle{Trends in Cognitive Sciences}}}  (\bibinfo{year}{2023}).

\bibitem{kidder2024empathy}
\bibinfo{author}{Kidder, W.}, \bibinfo{author}{D'Cruz, J.} \& \bibinfo{author}{Varshney, K.~R.}
\newblock \bibinfo{journal}{\bibinfo{title}{Empathy and the right to be an exception: What llms can and cannot do}}.
\newblock {\emph{\JournalTitle{arXiv preprint arXiv:2401.14523}}}  (\bibinfo{year}{2024}).

\bibitem{brown1987politeness}
\bibinfo{author}{Brown, P.} \& \bibinfo{author}{Levinson, S.~C.}
\newblock \emph{\bibinfo{title}{Politeness: Some universals in language usage}}.
\newblock \bibinfo{number}{4} (\bibinfo{publisher}{Cambridge university press}, \bibinfo{year}{1987}).

\bibitem{goffman1978presentation}
\bibinfo{author}{Goffman, E.}
\newblock \emph{\bibinfo{title}{The presentation of self in everyday life}} (\bibinfo{publisher}{Harmondsworth London}, \bibinfo{year}{1978}).

\bibitem{althoff2016large}
\bibinfo{author}{Althoff, T.}, \bibinfo{author}{Clark, K.} \& \bibinfo{author}{Leskovec, J.}
\newblock \bibinfo{journal}{\bibinfo{title}{Large-scale analysis of counseling conversations: An application of natural language processing to mental health}}.
\newblock {\emph{\JournalTitle{TACL}}}  (\bibinfo{year}{2016}).

\bibitem{de2017language}
\bibinfo{author}{De~Choudhury, M.} \& \bibinfo{author}{Kiciman, E.}
\newblock \bibinfo{title}{The language of social support in social media and its effect on suicidal ideation risk}.
\newblock In \emph{\bibinfo{booktitle}{Proceedings of the international AAAI conference on web and social media}}, vol.~\bibinfo{volume}{11}, \bibinfo{pages}{32--41} (\bibinfo{year}{2017}).

\bibitem{wang2021mutual}
\bibinfo{author}{Wang, Q.}, \bibinfo{author}{Saha, K.}, \bibinfo{author}{Gregori, E.}, \bibinfo{author}{Joyner, D.} \& \bibinfo{author}{Goel, A.}
\newblock \bibinfo{title}{Towards mutual theory of mind in human-ai interaction: How language reflects what students perceive about a virtual teaching assistant}.
\newblock In \emph{\bibinfo{booktitle}{Proceedings of the 2021 CHI conference on human factors in computing systems}}, \bibinfo{pages}{1--14} (\bibinfo{year}{2021}).

\bibitem{danescu2011mark}
\bibinfo{author}{Danescu-Niculescu-Mizil, C.}, \bibinfo{author}{Gamon, M.} \& \bibinfo{author}{Dumais, S.}
\newblock \bibinfo{title}{Mark my words! linguistic style accommodation in social media}.
\newblock In \emph{\bibinfo{booktitle}{Proceedings of the 20th international conference on World wide web}}, \bibinfo{pages}{745--754} (\bibinfo{year}{2011}).

\bibitem{cohen1985stress}
\bibinfo{author}{Cohen, S.} \& \bibinfo{author}{Wills, T.~A.}
\newblock \bibinfo{journal}{\bibinfo{title}{Stress, social support, and the buffering hypothesis.}}
\newblock {\emph{\JournalTitle{Psychological bulletin}}} \textbf{\bibinfo{volume}{98}}, \bibinfo{pages}{310} (\bibinfo{year}{1985}).

\bibitem{rains2009meta}
\bibinfo{author}{Rains, S.~A.} \& \bibinfo{author}{Young, V.}
\newblock \bibinfo{journal}{\bibinfo{title}{A meta-analysis of research on formal computer-mediated support groups: Examining group characteristics and health outcomes}}.
\newblock {\emph{\JournalTitle{Human communication research}}} \textbf{\bibinfo{volume}{35}}, \bibinfo{pages}{309--336} (\bibinfo{year}{2009}).

\bibitem{cutrona1992controllability}
\bibinfo{author}{Cutrona, C.~E.} \& \bibinfo{author}{Suhr, J.~A.}
\newblock \bibinfo{journal}{\bibinfo{title}{Controllability of stressful events and satisfaction with spouse support behaviors}}.
\newblock {\emph{\JournalTitle{Communication Research}}} \textbf{\bibinfo{volume}{19}}, \bibinfo{pages}{154--174} (\bibinfo{year}{1992}).

\bibitem{andalibi2018social}
\bibinfo{author}{Andalibi, N.}, \bibinfo{author}{Haimson, O.~L.}, \bibinfo{author}{Choudhury, M.~D.} \& \bibinfo{author}{Forte, A.}
\newblock \bibinfo{journal}{\bibinfo{title}{Social support, reciprocity, and anonymity in responses to sexual abuse disclosures on social media}}.
\newblock {\emph{\JournalTitle{ACM Transactions on Computer-Human Interaction (TOCHI)}}} \textbf{\bibinfo{volume}{25}}, \bibinfo{pages}{1--35} (\bibinfo{year}{2018}).

\bibitem{koko_gizmodo}
\bibinfo{title}{A mental health app tested chatgpt on its users. the founder said backlash was just a misunderstanding}.
\newblock \bibinfo{howpublished}{\url{https://gizmodo.com/mental-health-therapy-app-ai-koko-chatgpt-rob-morris-1849965534}} (\bibinfo{year}{2023}).
\newblock \bibinfo{note}{Accessed: 2024-03-10}.

\bibitem{koko_nbc}
\bibinfo{title}{A mental health tech company ran an ai experiment on real users. nothing’s stopping apps from conducting more}.
\newblock \bibinfo{howpublished}{\url{https://www.nbcnews.com/tech/internet/chatgpt-ai-experiment-mental-health-tech-app-koko-rcna65110}} (\bibinfo{year}{2023}).
\newblock \bibinfo{note}{Accessed: 2024-03-10}.

\bibitem{davis1983measuring}
\bibinfo{author}{Davis, M.~H.}
\newblock \bibinfo{journal}{\bibinfo{title}{Measuring individual differences in empathy: evidence for a multidimensional approach.}}
\newblock {\emph{\JournalTitle{Journal of personality and social psychology}}} \textbf{\bibinfo{volume}{44}}, \bibinfo{pages}{113} (\bibinfo{year}{1983}).

\bibitem{decety2004functional}
\bibinfo{author}{Decety, J.} \& \bibinfo{author}{Jackson, P.~L.}
\newblock \bibinfo{journal}{\bibinfo{title}{The functional architecture of human empathy}}.
\newblock {\emph{\JournalTitle{Behavioral and cognitive neuroscience reviews}}} \textbf{\bibinfo{volume}{3}}, \bibinfo{pages}{71--100} (\bibinfo{year}{2004}).

\bibitem{batson2009these}
\bibinfo{author}{Batson, C.~D.}
\newblock \bibinfo{journal}{\bibinfo{title}{These things called empathy: Eight related but distinct phenomena}}.
\newblock {\emph{\JournalTitle{The Social Neuroscience of Empathy}}} \bibinfo{pages}{3--16} (\bibinfo{year}{2009}).

\bibitem{bickmore2005establishing}
\bibinfo{author}{Bickmore, T.~W.} \& \bibinfo{author}{Picard, R.~W.}
\newblock \bibinfo{journal}{\bibinfo{title}{Establishing and maintaining long-term human-computer relationships}}.
\newblock {\emph{\JournalTitle{ACM Transactions on Computer-Human Interaction (TOCHI)}}} \textbf{\bibinfo{volume}{12}}, \bibinfo{pages}{293--327} (\bibinfo{year}{2005}).

\bibitem{dasswain2025ai}
\bibinfo{author}{Das~Swain, V.} \emph{et~al.}
\newblock \bibinfo{title}{Ai on my shoulder: Supporting emotional labor in front-office roles with an llm-based empathetic coworker}.
\newblock In \emph{\bibinfo{booktitle}{Proceedings of the 2025 CHI Conference on Human Factors in Computing Systems}} (\bibinfo{year}{2025}).

\bibitem{pennebaker2007expressive}
\bibinfo{author}{Pennebaker, J.~W.} \& \bibinfo{author}{Chung, C.~K.}
\newblock \bibinfo{journal}{\bibinfo{title}{Expressive writing, emotional upheavals, and health}}.
\newblock {\emph{\JournalTitle{Handbook of health psychology}}} \bibinfo{pages}{263--284} (\bibinfo{year}{2007}).

\bibitem{cheng2015antisocial}
\bibinfo{author}{Cheng, J.}, \bibinfo{author}{Danescu-Niculescu-Mizil, C.} \& \bibinfo{author}{Leskovec, J.}
\newblock \bibinfo{title}{Antisocial behavior in online discussion communities}.
\newblock In \emph{\bibinfo{booktitle}{International AAAI Conference on Web and Social Media}} (\bibinfo{year}{2015}).

\bibitem{tao2024cultural}
\bibinfo{author}{Tao, Y.}, \bibinfo{author}{Viberg, O.}, \bibinfo{author}{Baker, R.~S.} \& \bibinfo{author}{Kizilcec, R.~F.}
\newblock \bibinfo{journal}{\bibinfo{title}{Cultural bias and cultural alignment of large language models}}.
\newblock {\emph{\JournalTitle{PNAS nexus}}} \textbf{\bibinfo{volume}{3}}, \bibinfo{pages}{pgae346} (\bibinfo{year}{2024}).

\bibitem{shrawgi2024uncovering}
\bibinfo{author}{Shrawgi, H.}, \bibinfo{author}{Rath, P.}, \bibinfo{author}{Singhal, T.} \& \bibinfo{author}{Dandapat, S.}
\newblock \bibinfo{title}{Uncovering stereotypes in large language models: A task complexity-based approach}.
\newblock In \emph{\bibinfo{booktitle}{Proceedings of the 18th Conference of the European Chapter of the Association for Computational Linguistics (Volume 1: Long Papers)}}, \bibinfo{pages}{1841--1857} (\bibinfo{year}{2024}).

\bibitem{zhou2023synthetic}
\bibinfo{author}{Zhou, J.}, \bibinfo{author}{Zhang, Y.}, \bibinfo{author}{Luo, Q.}, \bibinfo{author}{Parker, A.~G.} \& \bibinfo{author}{De~Choudhury, M.}
\newblock \bibinfo{title}{Synthetic lies: Understanding ai-generated misinformation and evaluating algorithmic and human solutions}.
\newblock In \emph{\bibinfo{booktitle}{Proceedings of the 2023 CHI conference on human factors in computing systems}}, \bibinfo{pages}{1--20} (\bibinfo{year}{2023}).

\bibitem{jin2024better}
\bibinfo{author}{Jin, Y.} \emph{et~al.}
\newblock \bibinfo{title}{Better to ask in english: Cross-lingual evaluation of large language models for healthcare queries}.
\newblock In \emph{\bibinfo{booktitle}{Proceedings of the ACM Web Conference 2024}}, \bibinfo{pages}{2627--2638} (\bibinfo{year}{2024}).

\bibitem{chandra2024lived}
\bibinfo{author}{Chandra, M.} \emph{et~al.}
\newblock \bibinfo{title}{Lived experience not found: Llms struggle to align with experts on addressing adverse drug reactions from psychiatric medication use}.
\newblock In \emph{\bibinfo{booktitle}{2025 Annual Conference of the Nations of the Americas Chapter of the Association for Computational Linguistics (NAACL)}} (\bibinfo{year}{2025}).

\bibitem{weidinger2021ethical}
\bibinfo{author}{Weidinger, L.} \emph{et~al.}
\newblock \bibinfo{journal}{\bibinfo{title}{Ethical and social risks of harm from language models}}.
\newblock {\emph{\JournalTitle{arXiv preprint arXiv:2112.04359}}}  (\bibinfo{year}{2021}).

\bibitem{bender2021dangers}
\bibinfo{author}{Bender, E.~M.}, \bibinfo{author}{Gebru, T.}, \bibinfo{author}{McMillan-Major, A.} \& \bibinfo{author}{Shmitchell, S.}
\newblock \bibinfo{title}{On the dangers of stochastic parrots: Can language models be too big?}
\newblock In \emph{\bibinfo{booktitle}{Proceedings of the 2021 ACM conference on fairness, accountability, and transparency}}, \bibinfo{pages}{610--623} (\bibinfo{year}{2021}).

\bibitem{nori2023capabilities}
\bibinfo{author}{Nori, H.}, \bibinfo{author}{King, N.}, \bibinfo{author}{McKinney, S.~M.}, \bibinfo{author}{Carignan, D.} \& \bibinfo{author}{Horvitz, E.}
\newblock \bibinfo{journal}{\bibinfo{title}{Capabilities of gpt-4 on medical challenge problems}}.
\newblock {\emph{\JournalTitle{arXiv preprint arXiv:2303.13375}}}  (\bibinfo{year}{2023}).

\bibitem{timmons2023call}
\bibinfo{author}{Timmons, A.~C.} \emph{et~al.}
\newblock \bibinfo{journal}{\bibinfo{title}{A call to action on assessing and mitigating bias in artificial intelligence applications for mental health}}.
\newblock {\emph{\JournalTitle{Perspectives on Psychological Science}}} \textbf{\bibinfo{volume}{18}}, \bibinfo{pages}{1062--1096} (\bibinfo{year}{2023}).

\bibitem{sogancioglu2024fairness}
\bibinfo{author}{Sogancioglu, G.}, \bibinfo{author}{Mosteiro, P.}, \bibinfo{author}{Salah, A.~A.}, \bibinfo{author}{Scheepers, F.} \& \bibinfo{author}{Kaya, H.}
\newblock \bibinfo{title}{Fairness in ai-based mental health: Clinician perspectives and bias mitigation}.
\newblock In \emph{\bibinfo{booktitle}{Proceedings of the AAAI/ACM Conference on AI, Ethics, and Society}}, vol.~\bibinfo{volume}{7}, \bibinfo{pages}{1390--1400} (\bibinfo{year}{2024}).

\bibitem{kolla2024llm}
\bibinfo{author}{Kolla, M.}, \bibinfo{author}{Salunkhe, S.}, \bibinfo{author}{Chandrasekharan, E.} \& \bibinfo{author}{Saha, K.}
\newblock \bibinfo{title}{Llm-mod: Can large language models assist content moderation?}
\newblock In \emph{\bibinfo{booktitle}{Extended Abstracts of the CHI Conference on Human Factors in Computing Systems}}, \bibinfo{pages}{1--8} (\bibinfo{year}{2024}).

\bibitem{amershi2019guidelines}
\bibinfo{author}{Amershi, S.} \emph{et~al.}
\newblock \bibinfo{title}{Guidelines for human-ai interaction}.
\newblock In \emph{\bibinfo{booktitle}{Proceedings of the 2019 chi conference on human factors in computing systems}}, \bibinfo{pages}{1--13} (\bibinfo{year}{2019}).

\bibitem{ehsan2023charting}
\bibinfo{author}{Ehsan, U.}, \bibinfo{author}{Saha, K.}, \bibinfo{author}{De~Choudhury, M.} \& \bibinfo{author}{Riedl, M.~O.}
\newblock \bibinfo{journal}{\bibinfo{title}{Charting the sociotechnical gap in explainable ai: A framework to address the gap in xai}}.
\newblock {\emph{\JournalTitle{Proceedings of the ACM on human-computer interaction}}} \textbf{\bibinfo{volume}{7}}, \bibinfo{pages}{1--32} (\bibinfo{year}{2023}).

\bibitem{zhang2024dark}
\bibinfo{author}{Zhang, R.} \emph{et~al.}
\newblock \bibinfo{journal}{\bibinfo{title}{The dark side of ai companionship: A taxonomy of harmful algorithmic behaviors in human-ai relationships}}.
\newblock {\emph{\JournalTitle{arXiv preprint arXiv:2410.20130}}}  (\bibinfo{year}{2024}).

\bibitem{kim2023supporters}
\bibinfo{author}{Kim, M.}, \bibinfo{author}{Saha, K.}, \bibinfo{author}{De~Choudhury, M.} \& \bibinfo{author}{Choi, D.}
\newblock \bibinfo{journal}{\bibinfo{title}{Supporters first: understanding online social support on mental health from a supporter perspective}}.
\newblock {\emph{\JournalTitle{Proceedings of the ACM on Human-Computer Interaction}}} \textbf{\bibinfo{volume}{7}}, \bibinfo{pages}{1--28} (\bibinfo{year}{2023}).

\bibitem{saha2025ai}
\bibinfo{author}{Saha, K.}, \bibinfo{author}{Jain, Y.}, \bibinfo{author}{Liu, C.}, \bibinfo{author}{Kaliappan, S.} \& \bibinfo{author}{Karkar, R.}
\newblock \bibinfo{journal}{\bibinfo{title}{Ai vs. humans for online support: Comparing the language of responses from llms and online communities of alzheimer’s disease}}.
\newblock {\emph{\JournalTitle{ACM Transactions on Computing for Healthcare}}}  (\bibinfo{year}{2025}).

\bibitem{li2025emails}
\bibinfo{author}{Li, W.}, \bibinfo{author}{Lai, Y.}, \bibinfo{author}{Soni, S.} \& \bibinfo{author}{Saha, K.}
\newblock \bibinfo{title}{Emails by llms: A comparison of language in ai-generated and human-written emails}.
\newblock In \emph{\bibinfo{booktitle}{Proceedings of the 17th ACM Conference on Web Science}} (\bibinfo{year}{2025}).

\bibitem{yim2025generative}
\bibinfo{author}{Yim, S.~H.}, \bibinfo{author}{Yoo, D.~W.}, \bibinfo{author}{Polymerou, A.}, \bibinfo{author}{Liu, Y.} \& \bibinfo{author}{Saha, K.}
\newblock \bibinfo{journal}{\bibinfo{title}{Generative ai for eating disorders: linguistic comparison with online support and qualitative analysis of harms}}.
\newblock {\emph{\JournalTitle{International Journal of Eating Disorders}}}  (\bibinfo{year}{2025}).

\bibitem{schwartz2013personality}
\bibinfo{author}{Schwartz, H.~A.}, \bibinfo{author}{Eichstaedt, J.~C.}, \bibinfo{author}{Kern, M.~L.} \emph{et~al.}
\newblock \bibinfo{journal}{\bibinfo{title}{Personality, gender, and age in the language of social media: The open-vocabulary approach}}.
\newblock {\emph{\JournalTitle{PloS one}}} \textbf{\bibinfo{volume}{8}}, \bibinfo{pages}{e73791} (\bibinfo{year}{2013}).

\bibitem{tausczik2010psychological}
\bibinfo{author}{Tausczik, Y.~R.} \& \bibinfo{author}{Pennebaker, J.~W.}
\newblock \bibinfo{journal}{\bibinfo{title}{The psychological meaning of words: Liwc and computerized text analysis methods}}.
\newblock {\emph{\JournalTitle{Journal of language and social psychology}}} \textbf{\bibinfo{volume}{29}}, \bibinfo{pages}{24--54} (\bibinfo{year}{2010}).

\bibitem{coleman1975computer}
\bibinfo{author}{Coleman, M.} \& \bibinfo{author}{Liau, T.~L.}
\newblock \bibinfo{journal}{\bibinfo{title}{A computer readability formula designed for machine scoring.}}
\newblock {\emph{\JournalTitle{Journal of Applied Psychology}}} \textbf{\bibinfo{volume}{60}}, \bibinfo{pages}{283} (\bibinfo{year}{1975}).

\bibitem{pitler2008revisiting}
\bibinfo{author}{Pitler, E.} \& \bibinfo{author}{Nenkova, A.}
\newblock \bibinfo{title}{Revisiting readability: A unified framework for predicting text quality}.
\newblock In \emph{\bibinfo{booktitle}{Proceedings of the conference on empirical methods in natural language processing}}, \bibinfo{pages}{186--195} (\bibinfo{year}{2008}).

\bibitem{kolden2011congruence}
\bibinfo{author}{Kolden, G.~G.}, \bibinfo{author}{Klein, M.~H.}, \bibinfo{author}{Wang, C.-C.} \& \bibinfo{author}{Austin, S.~B.}
\newblock \bibinfo{journal}{\bibinfo{title}{Congruence/genuineness.}}
\newblock {\emph{\JournalTitle{Psychotherapy}}} \textbf{\bibinfo{volume}{48}}, \bibinfo{pages}{65} (\bibinfo{year}{2011}).

\bibitem{yuan2023mental}
\bibinfo{author}{Yuan, Y.}, \bibinfo{author}{Saha, K.}, \bibinfo{author}{Keller, B.}, \bibinfo{author}{Isomets{\"a}, E.~T.} \& \bibinfo{author}{Aledavood, T.}
\newblock \bibinfo{title}{Mental health coping stories on social media: A causal-inference study of papageno effect}.
\newblock In \emph{\bibinfo{booktitle}{Proceedings of the ACM Web Conference 2023}}, \bibinfo{pages}{2677--2685} (\bibinfo{year}{2023}).

\bibitem{heylighen1999formality}
\bibinfo{author}{Heylighen, F.} \& \bibinfo{author}{Dewaele, J.-M.}
\newblock \bibinfo{journal}{\bibinfo{title}{Formality of language: definition, measurement and behavioral determinants}}.
\newblock {\emph{\JournalTitle{Interner Bericht, Vrije Universiteit Brussel}}} \textbf{\bibinfo{volume}{4}} (\bibinfo{year}{1999}).

\bibitem{larsson2020syntactic}
\bibinfo{author}{Larsson, T.} \& \bibinfo{author}{Kaatari, H.}
\newblock \bibinfo{journal}{\bibinfo{title}{Syntactic complexity across registers: Investigating (in) formality in second-language writing}}.
\newblock {\emph{\JournalTitle{Journal of English for Academic Purposes}}} \textbf{\bibinfo{volume}{45}}, \bibinfo{pages}{100850} (\bibinfo{year}{2020}).

\bibitem{babakov2023don}
\bibinfo{author}{Babakov, N.}, \bibinfo{author}{Dale, D.}, \bibinfo{author}{Gusev, I.}, \bibinfo{author}{Krotova, I.} \& \bibinfo{author}{Panchenko, A.}
\newblock \bibinfo{title}{Don’t lose the message while paraphrasing: A study on content preserving style transfer}.
\newblock In \emph{\bibinfo{booktitle}{International Conference on Applications of Natural Language to Information Systems}} (\bibinfo{year}{2023}).

\bibitem{rao2018dear}
\bibinfo{author}{Rao, S.} \& \bibinfo{author}{Tetreault, J.}
\newblock \bibinfo{title}{Dear sir or madam, may i introduce the gyafc dataset: Corpus, benchmarks and metrics for formality style transfer}.
\newblock In \emph{\bibinfo{booktitle}{Proceedings of the 2018 Conference of the North American Chapter of the Association for Computational Linguistics: Human Language Technologies, Volume 1 (Long Papers)}}, \bibinfo{pages}{129--140} (\bibinfo{year}{2018}).

\bibitem{pavlick2016empirical}
\bibinfo{author}{Pavlick, E.} \& \bibinfo{author}{Tetreault, J.}
\newblock \bibinfo{journal}{\bibinfo{title}{An empirical analysis of formality in online communication}}.
\newblock {\emph{\JournalTitle{Transactions of the Association for Computational Linguistics}}} \textbf{\bibinfo{volume}{4}}, \bibinfo{pages}{61--74} (\bibinfo{year}{2016}).

\bibitem{sharma2023human}
\bibinfo{author}{Sharma, A.}, \bibinfo{author}{Lin, I.~W.}, \bibinfo{author}{Miner, A.~S.}, \bibinfo{author}{Atkins, D.~C.} \& \bibinfo{author}{Althoff, T.}
\newblock \bibinfo{journal}{\bibinfo{title}{Human--ai collaboration enables more empathic conversations in text-based peer-to-peer mental health support}}.
\newblock {\emph{\JournalTitle{Nature Machine Intelligence}}} \textbf{\bibinfo{volume}{5}}, \bibinfo{pages}{46--57} (\bibinfo{year}{2023}).

\bibitem{morris2018towards}
\bibinfo{author}{Morris, R.~R.}, \bibinfo{author}{Kouddous, K.}, \bibinfo{author}{Kshirsagar, R.} \& \bibinfo{author}{Schueller, S.~M.}
\newblock \bibinfo{journal}{\bibinfo{title}{Towards an artificially empathic conversational agent for mental health applications: system design and user perceptions}}.
\newblock {\emph{\JournalTitle{Journal of medical Internet research}}} \textbf{\bibinfo{volume}{20}}, \bibinfo{pages}{e10148} (\bibinfo{year}{2018}).

\bibitem{buechel2018modeling}
\bibinfo{author}{Buechel, S.}, \bibinfo{author}{Buffone, A.}, \bibinfo{author}{Slaff, B.}, \bibinfo{author}{Ungar, L.} \& \bibinfo{author}{Sedoc, J.}
\newblock \bibinfo{title}{Modeling empathy and distress in reaction to news stories}.
\newblock In \emph{\bibinfo{booktitle}{Proceedings of the 2018 Conference on Empirical Methods in Natural Language Processing}}, \bibinfo{pages}{4758--4765} (\bibinfo{year}{2018}).

\bibitem{tafreshi2021wassa}
\bibinfo{author}{Tafreshi, S.} \emph{et~al.}
\newblock \bibinfo{title}{Wassa 2021 shared task: predicting empathy and emotion in reaction to news stories}.
\newblock In \emph{\bibinfo{booktitle}{Workshop on Computational Approaches to Subjectivity and Sentiment Analysis (WASSA), held in conjunction with EACL 2021}}, \bibinfo{pages}{92--104} (\bibinfo{organization}{Association for Computational Linguistics}, \bibinfo{year}{2021}).

\bibitem{srinivasan2022tydip}
\bibinfo{author}{Srinivasan, A.} \& \bibinfo{author}{Choi, E.}
\newblock \bibinfo{title}{Tydip: A dataset for politeness classification in nine typologically diverse languages}.
\newblock In \emph{\bibinfo{booktitle}{Findings of the Association for Computational Linguistics: EMNLP 2022}}, \bibinfo{pages}{5723--5738} (\bibinfo{year}{2022}).

\bibitem{mikolov2013distributed}
\bibinfo{author}{Mikolov, T.}, \bibinfo{author}{Sutskever, I.}, \bibinfo{author}{Chen, K.}, \bibinfo{author}{Corrado, G.~S.} \& \bibinfo{author}{Dean, J.}
\newblock \bibinfo{title}{Distributed representations of words and phrases and their compositionality}.
\newblock In \emph{\bibinfo{booktitle}{Advances in neural information processing systems}}, \bibinfo{pages}{3111--3119} (\bibinfo{year}{2013}).

\bibitem{pennington2014glove}
\bibinfo{author}{Pennington, J.}, \bibinfo{author}{Socher, R.} \& \bibinfo{author}{Manning, C.}
\newblock \bibinfo{title}{Glove: Global vectors for word representation}.
\newblock In \emph{\bibinfo{booktitle}{Proceedings of the 2014 conference on empirical methods in natural language processing (EMNLP)}}, \bibinfo{pages}{1532--1543} (\bibinfo{year}{2014}).

\end{thebibliography}

\clearpage

\begin{table}[t!]
\centering
\sffamily
\footnotesize
\caption{Summary of comparison of \textbf{AI} (GPT-4-Turbo) and \textbf{OC} (Reddit) responses in terms of psycholinguistic categories as per Linguistic Inquiry and Word Count (LIWC~\cite{tausczik2010psychological}), along with Cohen's $d$, paired $t$-tests, and Kolmogorov-Smirnov (KS)-test. $p$-values reported after Bonferroni correction. (* $p$<0.05, ** $p$<0.01, *** $p$<0.001).}
\label{tab:liwc}
\begin{minipage}[t!]{0.49\columnwidth}
\centering
\resizebox{\columnwidth}{!}{
\begin{tabular}{lrrrrr@{}lr@{}l}
\setlength{\tabcolsep}{1pt}\\
\textbf{LIWC}&\textbf{AI}&\textbf{OC}&\textbf{$\Delta$\%}&\textbf{$d$} & \textbf{$t$}& &\textbf{KS}&\\ 
\toprule
\rowcollight \multicolumn{9}{l}{\textbf{Affect}}\\
Pos. Affect & 0.032 & 0.038 & -16.59 & -0.12 & -13.59 & *** & 0.28 & ***\\
Neg. Affect & 0.0 & 0.0 & -97.39 & -0.03 & -3.09 & ** & 0.0 & \\
Anxiety & 0.006 & 0.005 & 22.3 & 0.08 & 9.75 & *** & 0.19 & ***\\
\hdashline
Anger & 0.0 & 0.002 & -90.0 & \hlOC{-0.20} & -22.53 & *** & 0.07 & ***\\
Sad & 0.008 & 0.005 & 74.25 & \hlAI{0.24} & 30.03 & *** & 0.35 & ***\\
\rowcollight \multicolumn{9}{l}{\textbf{Cognition and Perception}}\\
Insight & 0.023 & 0.021 & 9.15 & 0.08 & 9.76 & *** & 0.33 & ***\\
Causation & 0.006 & 0.011 & -47.02 & \hlOC{-0.30} & -35.44 & *** & 0.13 & ***\\
Discrep. & 0.014 & 0.015 & -6.11 & -0.05 & -5.68 & *** & 0.36 & ***\\
\hdashline
Tentat. & 0.029 & 0.028 & 2.31 & 0.02 & 2.72 & ** & 0.29 & ***\\
Certainty & 0.005 & 0.012 & -58.0 & \hlOC{-0.29} & -32.11 & *** & 0.16 & ***\\
Differ. & 0.054 & 0.032 & 68.01 & \hlAI{0.63} & 88.92 & *** & 0.28 & ***\\
\hdashline
Percept & 0.017 & 0.017 & -1.12 & -0.01 & -0.94 &  & 0.32 & ***\\
See & 0.002 & 0.005 & -67.19 & \hlOC{-0.26} & -29.41 & *** & 0.1 & ***\\
Hear & 0.003 & 0.004 & -18.03 & -0.06 & -6.97 & *** & 0.15 & ***\\
\hdashline
Feel & 0.012 & 0.008 & 49.22 & \hlAI{0.23} & 27.91 & *** & 0.42 & ***\\
\rowcollight \multicolumn{9}{l}{\textbf{Social \& Personal Concerns}}\\
Family & 0.001 & 0.002 & -62.00 & -0.14 & -15.2 & *** & 0.04 & ***\\
Friend & 0.001 & 0.002 & -58.52 & -0.13 & -15.01 & *** & 0.05 & ***\\
Female & 0.002 & 0.004 & -58.43 & -0.17 & -20.56 & *** & 0.05 & ***\\
\hdashline
Male & 0.001 & 0.004 & -70.69 & \hlOC{-0.22} & -25.3 & *** & 0.06 & ***\\
Work & 0.006 & 0.006 & 2.48 & 0.01 & 1.19 &  & 0.2 & ***\\
Leisure & 0.001 & 0.002 & -49.01 & -0.10 & -10.9 & *** & 0.07 & ***\\
\hdashline
Home & 0.001 & 0.001 & -41.39 & -0.09 & -10.0 & *** & 0.06 & ***\\
Money & 0.001 & 0.001 & -31.97 & -0.06 & -7.13 & *** & 0.06 & ***\\
Religion & 0.0 & 0.001 & -92.23 & -0.10 & -10.7 & *** & 0.02 & ***\\
\hdashline
Death & 0.0 & 0.001 & -53.16 & -0.08 & -9.47 & *** & 0.01 & *\\
Motion & 0.005 & 0.011 & -50.36 & \hlOC{-0.31} & -35.13 & *** & 0.15 & ***\\
Space & 0.034 & 0.037 & -7.15 & -0.08 & -9.84 & *** & 0.22 & ***\\
\hdashline
Time & 0.011 & 0.029 & -60.17 & \hlOC{-0.57} & -66.03 & *** & 0.28 & ***\\
Affiliation & 0.016 & 0.013 & 25.04 & 0.13 & 15.59 & *** & 0.45 & ***\\
Achievement & 0.012 & 0.01 & 17.6 & 0.11 & 12.47 & *** & 0.4 & ***\\
\hdashline
Power & 0.031 & 0.011 & 192.6 & \hlAI{0.96} & 132.96 & *** & 0.52 & ***\\
Reward & 0.007 & 0.014 & -53.2 & \hlOC{-0.33} & -37.4 & *** & 0.15 & ***\\
Risk & 0.008 & 0.004 & 80.22 & \hlAI{0.29} & 36.32 & *** & 0.35 & ***\\
\bottomrule
\end{tabular}}
\end{minipage}\hfill
\begin{minipage}[t!]{0.49\columnwidth}
\centering
    \resizebox{\columnwidth}{!}{
\begin{tabular}{lrrrrr@{}lr@{}l}
\setlength{\tabcolsep}{0pt}\\
\textbf{LIWC} & \textbf{AI}  & \textbf{OC} & \textbf{$\Delta$\%} & \textbf{$d$} & \textbf{$t$}& & \textbf{KS} & \\ 
\toprule
\rowcollight \multicolumn{9}{l}{\textbf{Biological Processes}}\\
Body & 0.002 & 0.003 & -26.01 & -0.07 & -8.31 & *** & 0.1 & ***\\
Health & 0.019 & 0.006 & 206.96 & 0.68 & 90.23 & *** & 0.55 & ***\\
Sexual & 0.0 & 0.001 & -88.34 & -0.09 & -9.66 & *** & 0.01 & **\\
Ingest & 0.001 & 0.001 & -29.78 & -0.05 & -5.52 & *** & 0.07 & ***\\
\rowcollight \multicolumn{9}{l}{\textbf{Function Words}}\\
Article & 0.042 & 0.036 & 18.8 & \hlAI{0.23} & 27.89 & *** & 0.23 & ***\\
Preposition & 0.103 & 0.09 & 14.16 & \hlAI{0.27} & 33.79 & *** & 0.27 & ***\\
Aux. Verb & 0.094 & 0.074 & 26.40 & \hlAI{0.43} & 53.54 & *** & 0.34 & ***\\
\hdashline
Adverb & 0.044 & 0.052 & -16.57 & \hlOC{-0.21} & -24.77 & *** & 0.22 & ***\\
Conjunction & 0.066 & 0.056 & 17.69 & \hlAI{0.28} & 34.15 & *** & 0.25 & ***\\
Negation & 0.008 & 0.015 & -43.22 & \hlOC{-0.26} & -29.72 & *** & 0.16 & ***\\
\hdashline
Verb & 0.127 & 0.158 & -19.72 & \hlOC{-0.46} & -56.34 & *** & 0.3 & ***\\
Adjective & 0.053 & 0.041 & 27.44 & \hlAI{0.26} & 31.17 & *** & 0.31 & ***\\
Compare & 0.017 & 0.021 & -20.15 & -0.14 & -15.87 & *** & 0.36 & ***\\
\hdashline
Interrog. & 0.012 & 0.012 & -4.54 & -0.03 & -3.33 & *** & 0.38 & ***\\
Number & 0.001 & 0.004 & -67.65 & \hlOC{-0.28} & -31.81 & *** & 0.09 & ***\\
Quantifier & 0.009 & 0.018 & -50.38 & \hlOC{-0.40} & -46.03 & *** & 0.2 & ***\\
\rowcollight \multicolumn{9}{l}{\textbf{Interpersonal Focus (Pronouns)}}\\
1st P. Sin. & 0.017 & 0.058 & -70.1 & \hlOC{-0.93} & -107.9 & *** & 0.53 & ***\\
1st P. Plu. & 0.001 & 0.003 & -71.3 & \hlOC{-0.23} & -27.32 & *** & 0.1 & ***\\
2nd P. & 0.047 & 0.034 & 38.60 & \hlAI{0.32} & 37.73 & *** & 0.38 & ***\\
\hdashline
3rd P. Sin. & 0.003 & 0.006 & -56.77 & \hlOC{-0.22} & -26.48 & *** & 0.06 & ***\\
3rd P. Plu. & 0.005 & 0.005 & -1.53 & -0.01 & -0.75 &  & 0.25 & ***\\
Impersonal Prn. & 0.065 & 0.05 & 29.32 & \hlAI{0.37} & 45.87 & *** & 0.26 & ***\\
\rowcollight \multicolumn{9}{l}{\textbf{Temporal References}}\\
Past Focus & 0.013 & 0.035 & -61.9 & \hlOC{-0.48} & -53.59 & *** & 0.33 & ***\\
Present Focus & 0.126 & 0.109 & 15.41 & \hlAI{0.27} & 33.72 & *** & 0.28 & ***\\
Future Focus & 0.006 & 0.01 & -38.76 & \hlOC{-0.24} & -28.18 & *** & 0.15 & ***\\
\rowcollight \multicolumn{9}{l}{\textbf{Informal}}\\
Swear & 0.0 & 0.001 & -99.22 & -0.14 & -15.34 & *** & 0.05 & ***\\
Netspeak & 0.0 & 0.007 & -96.89 & \hlOC{-0.26} & -28.6 & *** & 0.16 & ***\\
Assent & 0.003 & 0.008 & -63.48 & -0.13 & -14.44 & *** & 0.17 & ***\\
\hdashline
Nonfluent & 0.0 & 0.002 & -80.11 & -0.18 & -19.56 & *** & 0.06 & ***\\
Filler & 0.0 & 0.0 & -100.0 & -0.07 & -8.08 & *** & 0.01 & **\\
\\
\bottomrule
\end{tabular}}
\end{minipage}
\end{table}

\clearpage
\begin{table}[t!]
\centering
\sffamily
\footnotesize
\caption{Summary of comparing the responses by \textcolor{aicolor}{\textbf{AI (GPT-4-Turbo)}} and by \textcolor{occolor}{\textbf{humans on online communities (Reddit)}} in terms of effect size (Cohen's $d$), paired $t$-test, and $KS$-test ($ * p <0.05, ** p<0.01, *** p<0.001$).}
\label{tab:lexicosemantics}
\begin{tabular}{lrrr@{}c@{}lrr@{}lr@{}l}
\setlength{\tabcolsep}{1pt}\\
\textbf{Categories} & \textbf{AI} & \textbf{Reddit} &\multicolumn{3}{c}{\textbf{Difference \%}}& \textbf{Cohen's $d$} & \textbf{$t$-test}& & \textbf{KS-test} & \\ 
\toprule
\rowcollight \multicolumn{11}{l}{\textbf{Linguistic Structure}}\\
Verbosity: Words \edit{Per Response}& 160.35 & 77.35 & & 107.30 &\aibarr{10.73} & 0.63 & 69.25 & *** & 0.35 & ***\\
Verbosity: Words Per Sentence & 19.28 & 13.76 & & 40.12&\aibarr{4.012} & 0.17 &  19.14 & *** & 0.34 & ***\\
Readability & 11.19 & 6.58 & & 70.06 &\aibarr{7.006} & 0.71 & 79.13 & *** & 0.64 & ***\\
Repeatability & 0.33 & 0.20 & & 66.56  &\aibarr{6.656} & 0.88 & 95.54 & *** & 0.40 & ***\\
Complexity & 4.63 & 3.31 & & 39.97 &\aibarr{3.997} & 0.75 & 83.52 & *** & 0.50 & ***\\
\rowcollight \multicolumn{11}{l}{\textbf{Linguistic Style}}\\
Categorical Dynamic Index (CDI) & 9.66 & 6.90 & & 40.04 &\aibarr{4.004} & 0.29  & 33.04 & *** & 0.25 & ***\\
Formality & 0.87 & 0.67 & & 30.14 & \aibarr{3.014} & 0.97 & 107.43 & *** & 0.51 & ***\\
Empathy & 0.84 & 0.71 & & 18.76 &\aibarr{1.876} & 0.63 & 69.19 & *** & 0.23 & ***\\
Politeness & 0.79 & 0.67 & & 17.99 & \aibarr{1.799} & 0.57 & 63.54 & *** & 0.26 & ***\\
\rowcollight \multicolumn{11}{l}{\textbf{Adaptability to Query}}\\
Semantic Similarity & 0.71 & 0.59 & & 21.26 &\aibarr{2.126} & 0.52  & 63.13 & *** & 0.30 & ***\\
Linguistic Style Accommodation & 0.77 & 0.71 & & 8.83 & \aibarr{0.883} & 0.21  & 23.28 & *** & 0.22 & ***\\
Diversity/Creativity & 0.06 & 0.13 & \ocbarr{5.704}&-57.04 & & -0.89 & -103.37 & *** & 0.46 & ***\\
\rowcollight\multicolumn{11}{l}{Social Support}\\
Emotional Support & 0.79 & 0.49 & & 62.43 &\aibarr{6.243}  & 0.78  & 89.90 & *** & 0.57 & ***\\
Informational Support & 0.62 & 0.52 & & 19.90 &\aibarr{1.990}  & 0.24  & 25.69 & *** & 0.36 & ***\\
\bottomrule
\end{tabular}
\end{table}

\newpage

\begin{table}[t]
\centering
\sffamily
\footnotesize
{\color{editCol}
\caption{\edit{Summary of expert clinical ratings for AI-generated responses (N=100). Ratings were on a 1--5 scale, with higher values indicating stronger performance: greater accuracy, safety, emotional attunement, and contextual responsiveness.}}
\label{tab:expert_ratings}
\begin{tabular}{lcccc}
\textbf{Dimension} & \textbf{N} & \textbf{Mean} & \textbf{SD} & \textbf{\% $\geq$ 4} \\
\toprule
Factual Accuracy                      & 100 & 5.00 & 0.00 & 100\% \\
Potential Harmfulness               & 100 & 1.14 & 0.35 & 0\% \\
Emotional Attunement (Empathy)         & 100 & 1.82 & 0.99 & 0\% \\
Contextual Responsiveness (Addressing) & 100 & 2.64 & 0.67 & 0\% \\
\bottomrule
\end{tabular}
}
\end{table}

\clearpage
\begin{table}[t!]
\centering
\sffamily
\footnotesize
\caption{Summary of comparison of \textbf{Reddit} and \textbf{AI} (for GPT-4-Turbo, Llama-3, and Mistral-7B) responses in terms of psycholinguistic categories as per Linguistic Inquiry and Word Count (LIWC~\cite{tausczik2010psychological}), along with Cohen's $d$, paired $t$-tests, and Kolmogorov-Smirnov (KS)-test. $p$-values reported after Bonferroni correction. (* $p$<0.05, ** $p$<0.01, *** $p$<0.001).}
\label{tab:liwc_llms}
\begin{minipage}[t!]{0.5\columnwidth}
\centering
\resizebox{\columnwidth}{!}{
\begin{tabular}{lrrr@{}lrr@{}lrr@{}lr@{}l}
\setlength{\tabcolsep}{0pt}\\
& \textbf{Reddit} & \multicolumn{3}{c}{\textbf{GPT}}& \multicolumn{3}{c}{\textbf{Llama}}& \multicolumn{3}{c}{\textbf{Mistral}} & & \\ 
\cmidrule(lr){2-2}\cmidrule(lr){3-5}\cmidrule(lr){6-8}\cmidrule(lr){9-11}
\textbf{Categories}  & \textbf{Mean} & \textbf{Mean} &  \textbf{t-test} &  & \textbf{Mean} & \textbf{t-test} &  & \textbf{Mean} & \textbf{t-test} &  & \textbf{H-stat.}& \\ 
\toprule
\rowcollight \multicolumn{13}{l}{\textbf{Affect}}\\
Pos. Affect & 0.038 & 0.032 & -13.59 & *** & 0.027 & -25.2 & *** & 0.035 & -6.29 & *** & 7863.22 & ***\\
Neg. Affect & 0.0 & 0.0 & -3.09 & *** & 0.0 & -2.19 & *** & 0.0 & -2.67 & *** & 43.21 & ***\\
Anxiety & 0.005 & 0.006 & 9.75 & *** & 0.007 & 20.07 & *** & 0.007 & 21.73 & *** & 24817.71 & ***\\
\hdashline
Anger & 0.002 & 0.0 & -22.53 & *** & 0.0 & -20.47 & *** & 0.0 & -20.98 & *** & 1250.36 & ***\\
Sad & 0.005 & 0.008 & 30.03 & *** & 0.003 & -17.63 & *** & 0.004 & -10.42 & *** & 19752.18 & ***\\
\rowcollight \multicolumn{13}{l}{\textbf{Cognition and Perception}}\\
Insight & 0.021 & 0.023 & 9.76 & *** & 0.024 & 18.23 & *** & 0.029 & 43.76 & *** & 14517.03 & ***\\
Causation & 0.011 & 0.006 & -35.44 & *** & 0.008 & -21.88 & *** & 0.009 & -14.44 & *** & 7082.33 & ***\\
Discrep. & 0.015 & 0.014 & -5.68 & *** & 0.009 & -34.87 & *** & 0.013 & -13.17 & *** & 5585.53 & ***\\
\hdashline
Tentat. & 0.028 & 0.029 & 2.72 & *** & 0.032 & 15.45 & *** & 0.039 & 47.71 & *** & 10975.77 & ***\\
Certainty & 0.012 & 0.005 & -32.11 & *** & 0.007 & -21.57 & *** & 0.007 & -20.84 & *** & 7113.61 & ***\\
Differ. & 0.032 & 0.054 & 88.92 & *** & 0.033 & 6.29 & *** & 0.036 & 17.36 & *** & 10695.8 & ***\\
\hdashline
Percept & 0.017 & 0.017 & -0.94 & *** & 0.014 & -13.58 & *** & 0.015 & -7.13 & *** & 7416.95 & ***\\
See & 0.005 & 0.002 & -29.41 & *** & 0.002 & -21.77 & *** & 0.002 & -24.31 & *** & 6858.96 & ***\\
Hear & 0.004 & 0.003 & -6.97 & *** & 0.002 & -19.07 & *** & 0.003 & -4.15 & *** & 19374.27 & ***\\
\hdashline
Feel & 0.008 & 0.012 & 27.91 & *** & 0.009 & 6.66 & *** & 0.009 & 10.23 & *** & 25835.34 & ***\\
\rowcollight \multicolumn{9}{l}{\textbf{Social \& Personal Concerns}}\\
Family & 0.002 & 0.001 & -15.2 & *** & 0.002 & -1.78 & *** & 0.002 & -3.52 & *** & 24678.24 & ***\\
Friend & 0.002 & 0.001 & -15.01 & *** & 0.002 & -1.17 & *** & 0.002 & -0.93 & *** & 21835.96 & ***\\
Female & 0.004 & 0.002 & -20.56 & *** & 0.002 & -18.46 & *** & 0.002 & -18.42 & *** & 583.94 & ***\\
\hdashline
Male & 0.004 & 0.001 & -25.3 & *** & 0.002 & -23.59 & *** & 0.001 & -26.13 & *** & 840.37 & ***\\
Work & 0.006 & 0.006 & 1.19 & *** & 0.01 & 36.34 & *** & 0.011 & 48.55 & *** & 40897.5 & ***\\
Leisure & 0.002 & 0.001 & -10.9 & *** & 0.002 & 1.69 & *** & 0.002 & -0.78 & *** & 32357.44 & ***\\
\hdashline
Home & 0.001 & 0.001 & -10.0 & *** & 0.002 & 15.8 & *** & 0.002 & 8.96 & *** & 40369.15 & ***\\
Money & 0.001 & 0.001 & -7.13 & *** & 0.001 & 0.57 & *** & 0.001 & 3.82 & *** & 16647.94 & ***\\
Religion & 0.001 & 0.0 & -10.7 & *** & 0.0 & -10.25 & *** & 0.0 & -10.64 & *** & 587.1 & ***\\
\hdashline
Death & 0.001 & 0.0 & -9.47 & *** & 0.0 & -7.8 & *** & 0.0 & -6.73 & *** & 574.1 & ***\\
Relativity & 0.076 & 0.051 & -57.97 & *** & 0.053 & -53.66 & *** & 0.057 & -43.02 & *** & 6968.49 & ***\\
Motion & 0.011 & 0.005 & -35.13 & *** & 0.009 & -13.43 & *** & 0.009 & -10.49 & *** & 9868.06 & ***\\
\hdashline
Space & 0.037 & 0.034 & -9.84 & *** & 0.03 & -24.82 & *** & 0.033 & -12.44 & *** & 521.14 & ***\\
Time & 0.029 & 0.011 & -66.03 & *** & 0.014 & -56.99 & *** & 0.015 & -52.44 & *** & 4374.67 & ***\\
Drives & 0.046 & 0.06 & 39.31 & *** & 0.046 & -0.32 & *** & 0.056 & 29.39 & *** & 11091.23 & ***\\
\hdashline
Affiliation & 0.013 & 0.016 & 15.59 & *** & 0.017 & 17.52 & *** & 0.019 & 27.18 & *** & 29398.41 & ***\\
Achievement & 0.01 & 0.012 & 12.47 & *** & 0.011 & 7.97 & *** & 0.013 & 17.55 & *** & 21406.99 & ***\\
Power & 0.011 & 0.031 & 132.96 & *** & 0.017 & 41.61 & *** & 0.023 & 86.98 & *** & 52883.29 & ***\\
\hdashline
Reward & 0.014 & 0.007 & -37.4 & *** & 0.009 & -26.83 & *** & 0.01 & -20.38 & *** & 5844.18 & ***\\
Risk & 0.004 & 0.008 & 36.32 & *** & 0.004 & 2.21 & *** & 0.004 & 2.04 & *** & 23790.9 & ***\\
\bottomrule
\end{tabular}}
\end{minipage}\hfill
\begin{minipage}[t!]{0.5\columnwidth}
\centering
    \resizebox{\columnwidth}{!}{
\begin{tabular}{lrrr@{}lrr@{}lrr@{}lr@{}l}
\setlength{\tabcolsep}{0pt}\\
& \textbf{Reddit} & \multicolumn{3}{c}{\textbf{GPT}}& \multicolumn{3}{c}{\textbf{Llama}}& \multicolumn{3}{c}{\textbf{Mistral}} & & \\ 
\cmidrule(lr){2-2}\cmidrule(lr){3-5}\cmidrule(lr){6-8}\cmidrule(lr){9-11}
\textbf{Categories}  & \textbf{Mean} & \textbf{Mean} &  \textbf{t-test} &  & \textbf{Mean} & \textbf{t-test} &  & \textbf{Mean} & \textbf{t-test} &  & \textbf{H-stat.}& \\ 
\toprule
\rowcollight \multicolumn{13}{l}{\textbf{Biological Processes}}\\
Body & 0.003 & 0.002 & -8.31 & *** & 0.002 & -2.62 & *** & 0.003 & -1.2 & *** & 13152.44 & ***\\
Health & 0.006 & 0.019 & 90.23 & *** & 0.008 & 16.76 & *** & 0.01 & 29.89 & *** & 47840.27 & ***\\
Sexual & 0.001 & 0.0 & -9.66 & *** & 0.0 & -8.65 & *** & 0.0 & -9.27 & *** & 213.35 & ***\\
\hdashline
Ingest & 0.001 & 0.001 & -5.52 & *** & 0.001 & -4.81 & *** & 0.001 & 3.51 & *** & 14768.61 & ***\\
\rowcollight \multicolumn{13}{l}{\textbf{Function Words}}\\
Article & 0.036 & 0.042 & 27.89 & *** & 0.038 & 10.12 & *** & 0.04 & 18.6 & *** & 5024.7 & ***\\
Preposition & 0.09 & 0.103 & 33.79 & *** & 0.096 & 14.94 & *** & 0.111 & 54.96 & *** & 7867.38 & ***\\
Aux. Verb & 0.074 & 0.094 & 53.54 & *** & 0.068 & -16.76 & *** & 0.08 & 14.64 & *** & 10565.96 & ***\\
\hdashline
Adverb & 0.052 & 0.044 & -24.77 & *** & 0.026 & -77.28 & *** & 0.029 & -68.74 & *** & 13029.35 & ***\\
Conjunction & 0.056 & 0.066 & 34.15 & *** & 0.063 & 24.61 & *** & 0.075 & 66.18 & *** & 12180.34 & ***\\
Negation & 0.015 & 0.008 & -29.72 & *** & 0.01 & -23.09 & *** & 0.008 & -29.98 & *** & 2694.12 & ***\\
\hdashline
Verb & 0.158 & 0.127 & -56.34 & *** & 0.109 & -90.36 & *** & 0.133 & -46.08 & *** & 17285.76 & ***\\
Adjective & 0.041 & 0.053 & 31.17 & *** & 0.042 & 2.98 & *** & 0.047 & 15.82 & *** & 11915.15 & ***\\
Compare & 0.021 & 0.017 & -15.87 & *** & 0.015 & -23.7 & *** & 0.016 & -19.38 & *** & 2140.44 & ***\\
\hdashline
Interrog. & 0.012 & 0.012 & -3.33 & *** & 0.009 & -19.13 & *** & 0.008 & -27.73 & *** & 8116.9 & ***\\
Number & 0.004 & 0.001 & -31.81 & *** & 0.002 & -25.11 & *** & 0.002 & -20.11 & *** & 8637.13 & ***\\
Quantifier & 0.018 & 0.009 & -46.03 & *** & 0.01 & -39.79 & *** & 0.013 & -25.59 & *** & 2721.46 & ***\\
\rowcollight \multicolumn{13}{l}{\textbf{Interpersonal Focus (Pronouns)}}\\
1st P. Sin. & 0.058 & 0.017 & -107.9 & *** & 0.007 & -135.27 & *** & 0.008 & -134.79 & *** & 31931.33 & ***\\
1st P. Plu. & 0.003 & 0.001 & -27.32 & *** & 0.001 & -25.27 & *** & 0.001 & -22.31 & *** & 1831.54 & ***\\
2nd P. & 0.034 & 0.047 & 37.73 & *** & 0.052 & 52.42 & *** & 0.052 & 53.81 & *** & 26344.5 & ***\\
\hdashline
3rd P. Sin. & 0.006 & 0.003 & -26.48 & *** & 0.003 & -25.27 & *** & 0.003 & -26.86 & *** & 650.38 & ***\\
3rd P. Plu. & 0.005 & 0.005 & -0.75 & *** & 0.004 & -8.43 & *** & 0.005 & 6.27 & *** & 28430.59 & ***\\
Impersonal Prn. & 0.05 & 0.065 & 45.87 & *** & 0.041 & -28.58 & *** & 0.05 & 0.4 & *** & 8796.2 & ***\\
\rowcollight \multicolumn{13}{l}{\textbf{Temporal References}}\\
Past Focus & 0.035 & 0.013 & -53.59 & *** & 0.01 & -61.2 & *** & 0.013 & -54.44 & *** & 1032.16 & ***\\
Present Focus & 0.109 & 0.126 & 33.72 & *** & 0.097 & -23.47 & *** & 0.114 & 10.55 & *** & 7961.49 & ***\\
Future Focus & 0.01 & 0.006 & -28.18 & *** & 0.007 & -19.38 & *** & 0.011 & 4.25 & *** & 14879.24 & ***\\
\rowcollight \multicolumn{13}{l}{\textbf{Informal}}\\
Swear & 0.001 & 0.0 & -15.34 & *** & 0.0 & -14.98 & *** & 0.0 & -15.34 & *** & 2971.87 & ***\\
Netspeak & 0.007 & 0.0 & -28.6 & *** & 0.0 & -28.12 & *** & 0.0 & -28.68 & *** & 8091.97 & ***\\
Assent & 0.008 & 0.003 & -14.44 & *** & 0.002 & -17.82 & *** & 0.002 & -17.62 & *** & 10680.26 & ***\\
\hdashline
Nonfluent & 0.002 & 0.0 & -19.56 & *** & 0.001 & -14.34 & *** & 0.002 & -8.62 & *** & 13688.5 & ***\\
Filler & 0.0 & 0.0 & -8.08 & *** & 0.0 & -8.09 & *** & 0.0 & -8.07 & *** & 834.13 & ***\\
\\
\\
\bottomrule
\end{tabular}}
\end{minipage}
\end{table}

\clearpage
\begin{table}[t!]
\centering
\sffamily
\footnotesize
\caption{Summary of comparing the responses on Reddit and by multiple LLMs: GPT-4-Turbo, Llama-3.1, and Mistral-7B, including paired $t$-tests in comparison with Reddit responses, and a Kruskal-Wallis $H$-test across all the four modalities---Reddit, GPT, Llama, and Mistral ($ * p <0.05, ** p<0.01, *** p<0.001$).}
\label{tab:lexicosemantics_llms}
\begin{tabular}{lrrr@{}lrr@{}lrr@{}lr@{}l}
\setlength{\tabcolsep}{1pt}\\
& \textbf{Reddit} & \multicolumn{3}{c}{\textbf{GPT}}& \multicolumn{3}{c}{\textbf{Llama}}& \multicolumn{3}{c}{\textbf{Mistral}} & & \\ 
\cmidrule(lr){2-2}\cmidrule(lr){3-5}\cmidrule(lr){6-8}\cmidrule(lr){9-11}
\textbf{Categories}  & \textbf{Mean} & \textbf{Mean} &  \textbf{t-test} &  & \textbf{Mean} & \textbf{t-test} &  & \textbf{Mean} & \textbf{t-test} &  & \textbf{H-stat.}& \\ 
\toprule
\rowcollight \multicolumn{13}{l}{\textbf{Verbosity}}\\
Words \edit{Per Response}& 77.35 & 160.35 & 69.25 & *** & 470.55 & 306.3 & *** & 332.03 & 308.89 & *** & 51336.42 & ***\\
Words Per Sentence & 13.76 & 19.28 & 19.14 & *** & 18.19 & 39.86 & *** & 20.11 & 95.78 & *** & 14141.34 & ***\\
\rowcollight \multicolumn{13}{l}{\textbf{Linguistic Structure}}\\
Readability & 6.58 & 11.18 & 79.13 & *** & 10.11 & 49.99 & *** & 10.1 & 49.75 & *** & 24012.01 & ***\\
Repeatability & 0.2 & 0.33 & 95.54 & *** & 0.53 & 269.3 & *** & 0.48 & 277.59 & *** & 50929.85 & ***\\
Complexity & 3.31 & 4.63 & 83.52 & *** & 4.17 & 88.87 & *** & 4.39 & 110.81 & *** & 26327.51 & ***\\
\rowcollight \multicolumn{13}{l}{\textbf{Linguistic Style}}\\
Categorical Dynamic Index (CDI) & 6.9 & 9.66 & 33.04 & *** & 15.87 & 117.63 & *** & 14.0 & 97.92 & *** & 17245.87 & ***\\
Formality & 0.67 & 0.87 & 107.43 & *** & 0.03 & -299.46 & *** & 0.04 & -292.68 & *** & 72616.46 & ***\\
Empathy & 0.71 & 0.84 & 69.19 & *** & 0.03 & -307.72 & *** & 0.03 & -307.23 & *** & 67449.95 & ***\\
Politeness & 0.67 & 0.79 & 63.52 & *** & 0.03 & -307.47 & *** & 0.03 & -308.03 & *** & 68856.98 & ***\\
\rowcollight \multicolumn{13}{l}{\textbf{Adaptability to Query}}\\
Semantic Similarity & 0.59 & 0.71 & 63.13 & *** & 0.75 & 83.13 & *** & 0.76 & 87.94 & *** & 7331.63 & ***\\
Linguistic Style Accommodation & 0.71 & 0.77 & 23.28 & *** & 0.81 & 45.56 & *** & 0.83 & 51.44 & *** & 3731.16 & ***\\
Diversity/Creativity & 0.13 & 0.06 & -103.37 & *** & 0.09 & -35.11 & *** & 0.08 & -44.79 & *** & 13080.62 & ***\\
\rowcollight\multicolumn{13}{l}{\textbf{Social Support}}\\
Emotional Support & 0.49 & 0.79 & 89.9 & *** & 0.82 & 101.99 & *** & 0.86 & 124.4 & *** & 21035.19 & ***\\
Informational Support & 0.52 & 0.62 & 25.69 & *** & 0.94 & 150.25 & *** & 0.96 & 166.63 & *** & 30201.57 & ***\\
\bottomrule
\end{tabular}
\end{table}
\end{document}